\documentclass[10pt,twocolumn,letterpaper]{article}

\usepackage[pagenumbers]{cvpr} %

\usepackage{soul}
\usepackage{multirow}
\usepackage{xcolor}
\usepackage{xspace}
\newcommand{\whitetxt}[1]{{\color{white}#1}\normalfont}
\usepackage{color}
\usepackage[dvipsnames]{xcolor}
\usepackage{subcaption}
\usepackage{epsfig}
\usepackage{graphicx,animate}
\usepackage{tabularx}

\newbox\jsavebox

\definecolor{overviewred}{rgb}{1.0, 0.19, 0.19}
\definecolor{blue(munsell)}{rgb}{0.0, 0.5, 0.69}

\definecolor{R1color}{RGB}{0, 0, 255}      %
\definecolor{R2color}{RGB}{0, 176, 240}      %
\definecolor{R3color}{RGB}{0, 255, 0}       %

\definecolor{cvprblue}{rgb}{0.21,0.49,0.74}
\usepackage[pagebackref,breaklinks,colorlinks,allcolors=cvprblue]{hyperref}

\def\paperID{} %
\def\confName{CVPR}
\def\confYear{2026}

\title{Let it Snow! Animating 3D Gaussian Scenes with Dynamic Weather Effects \\ via Physics-Guided Score Distillation}

\author{
Gal Fiebelman$^{1}$ \ \ \
Hadar Averbuch-Elor$^{2}$ \ \ \
Sagie Benaim$^{1}$
\\[2mm]
\vspace{1em}
$^1$The Hebrew University of Jerusalem \ \ \
$^2$Cornell University
\\
\small{\url{https://galfiebelman.github.io/let-it-snow/}}
}

\begin{document}
\twocolumn[{
\renewcommand\twocolumn[1][]{#1}%
\maketitle
\begin{center}
    \centering
    \captionsetup{type=figure}
    \includegraphics[width=0.99\textwidth]{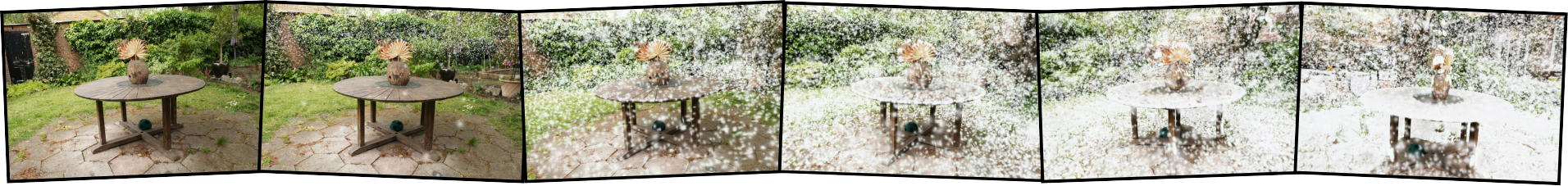}
    \includegraphics[width=0.99\textwidth]{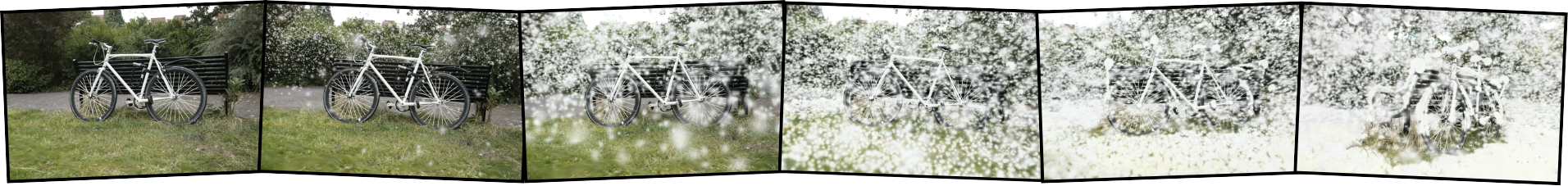}
    \includegraphics[width=0.99\textwidth]{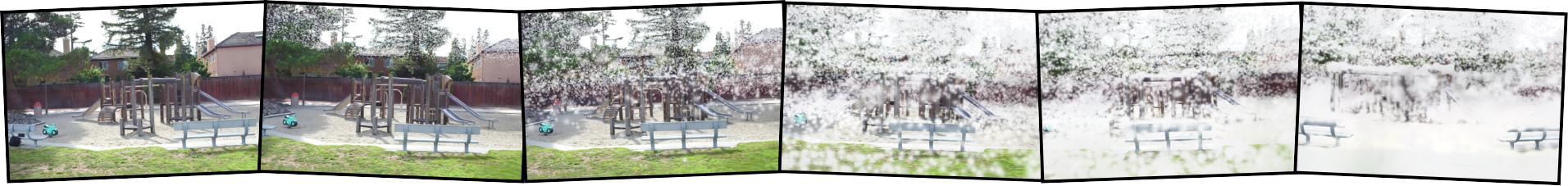}
    \vspace{-2mm}
    \caption{
    Our \textit{Physics-Guided Score Distillation} framework enables dynamic scene-wide editing of static 3D Gaussian scenes introducing dynamic weather effects such as snowfall, rainfall, fog, and sandstorms. We do so by introducing a unified optimization framework where physics simulation guides Score Distillation to jointly refine the motion prior for photorealism while simultaneously optimizing appearance. 
    Above, we show the effect of snow falling across different viewpoints at increasing timesteps over three different scenes.}

\label{fig:teaser}
\end{center}
}]

\begin{abstract}
3D Gaussian Splatting has recently enabled fast and photorealistic reconstruction of static 3D scenes. However, dynamic editing of such scenes remains a significant challenge. 
We introduce a novel framework, \textit{Physics-Guided Score Distillation}, to address a fundamental conflict: physics simulation provides a strong motion prior that is insufficient for photorealism, while video-based Score Distillation Sampling (SDS) alone cannot generate coherent motion for complex, multi-particle scenarios. We resolve this through a unified optimization framework where physics simulation guides Score Distillation to jointly refine the motion prior for photorealism while simultaneously optimizing appearance. Specifically, we learn a neural dynamics model that predicts particle motion and appearance, optimized end-to-end via a combined loss integrating Video-SDS for photorealism with our physics-guidance prior. This allows for photorealistic refinements while ensuring the dynamics remain plausible. Our framework enables scene-wide dynamic weather effects, including snowfall, rainfall, fog, and sandstorms, with physically plausible motion. Experiments demonstrate our physics-guided approach significantly outperforms baselines, with ablations confirming this joint refinement is essential for generating coherent, high-fidelity dynamics.

\end{abstract}
    
\section{Introduction}

\label{sec:intro}
Dynamic editing of static 3D scenes is typically a manual task that requires extensive knowledge of specialized software. A key challenge lies in creating temporally consistent, scene-wide modifications while maintaining both physical plausibility and photorealistic appearance. While static editing methods can enhance visual appearance, introducing dynamic elements that evolve over time and interact naturally across an entire scene remains largely unexplored. Recent advances in neural scene representations, particularly 3D Gaussian Splatting (3DGS)~\cite{kerbl20233d}, have enabled real-time, high-fidelity reconstruction and rendering of static 3D scenes from multi-view images. However, the development of effective approaches for dynamically editing these representations with scene-wide temporal effects presents a significant challenge.
    
Prior works for static editing of neural representations~\cite{li2023climatenerf, haque2023instruct, zhuang2023dreameditor, sella2023vox, chen2024gaussianeditor, wang2024gaussianeditor, wu2024gaussctrl} cannot model temporal evolution or dynamic effects. Physics-based animation approaches~\cite{xie2024physgaussian, li2023pac, huang2024dreamphysics, zhang2024physdreamer, qiu2024feature} provide plausible motion but focus on manipulating existing scene elements; they cannot synthesize photorealistic appearance for new elements, and their motion-only guidance is insufficient for photorealism. Conversely, data-driven 4D generation methods~\cite{ren2023dreamgaussian4d, jiang2025animate3d, zhao2023animate124, ling2024align, wimmer2024gaussians, mou2024instruct, kwon2025efficient} generate photorealistic appearance but struggle with physically plausible motion, often failing in scene-wide, multi-particle scenarios that require continuous particle emission.

Accordingly, we propose a framework for scene-wide, multi-particle dynamic editing that resolves this conflict through \emph{Physics-Guided Score Distillation}. Our key insight is that physics simulation can serve as a powerful motion prior to guide video diffusion optimization, enabling Score Distillation Sampling (SDS) to jointly refine motion for photorealism while synthesizing appearance, rather than learning both from scratch, a task that proves intractable for complex multi-particle scenarios. Rather than rigidly separating motion generation from appearance optimization, we jointly optimize both within a unified framework. 

Specifically, we learn a neural dynamics model that predicts particle trajectories and appearance parameters over time, supporting continuous particle emission and interaction throughout the temporal sequence. This model is optimized end-to-end using Video Score Distillation Sampling (Video-SDS), where physics guides the optimization through two complementary mechanisms: (1) the network is conditioned on reference trajectories from Material Point Method (MPM) simulation, providing physically-informed input features, and (2) a physics regularization loss encourages learned motion to remain close to simulated trajectories while permitting learned refinements for photorealism.

Our approach offers several advantages over existing editing methods. By leveraging physics as a guidance prior rather than a fixed constraint, we achieve temporally consistent edits with natural interaction patterns while allowing the diffusion model to refine both motion and appearance for visual quality. Our Video-SDS optimization, guided by physics regularization, ensures that dynamic elements achieve both photorealistic appearance and plausible motion that integrate seamlessly with the static scene.

We demonstrate our approach on scenes from both the MipNeRF 360~\cite{barron2022mip} and Tanks and Temples~\cite{knapitsch2017tanks} datasets, showing its effectiveness across a range of dynamic scene modifications, including snowfall, rainfall, fog, and sandstorms with diverse appearance variations. Extensive experiments demonstrate that our physics-guided approach significantly outperforms existing methods in both motion plausibility and visual quality. Our ablations confirm that when motion is fixed without physics-guided refinement, Video-SDS cannot properly optimize appearance demonstrating that motion refinement through our joint optimization is essential for achieving photorealistic results. Without physics guidance, Video-SDS alone cannot reliably learn both motion and appearance for complex multi-particle scenarios with continuous particle emission.
To summarize, our main contributions are:

\begin{itemize}

\item A \emph{Physics-Guided Score Distillation} framework that leverages physics simulation as a motion prior to guide video diffusion optimization, enabling joint learning of motion and appearance for scene-wide, multi-particle dynamic editing.

\item A neural dynamics model optimized through a combined loss integrating Video-SDS for photorealism with physics regularization for motion plausibility, supporting continuous particle emission over time.

\item State-of-the-art results across diverse dynamic weather effects, with ablations confirming that physics guidance is essential for coherent multi-particle dynamics.

\end{itemize}

\begin{figure*}
    \centering
    \includegraphics[width=\textwidth]{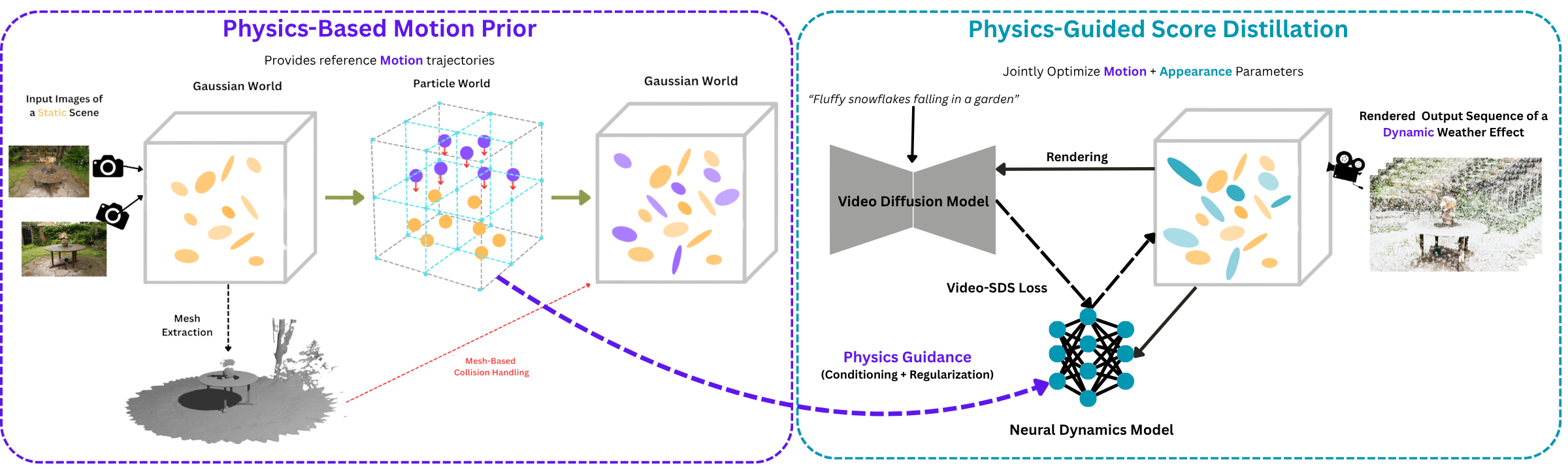}
    \vspace{-4mm}

    \caption{\textbf{Overview}. 
    \textcolor {Purple}{Left: Physics-Based Motion Prior.} Given multi-view images of a \textcolor{Peach}{static} scene, we reconstruct the scene using 3DGS, map the \textcolor{Peach}{static} Gaussians to a particle representation, introduce \textcolor{Purple}{dynamic} particles, and simulate their motion using the Material Point Method (MPM). We then map these particles back to the Gaussian world and refine scene interactions with our \textcolor{overviewred}{mesh-based collision handling} technique. The simulation provides reference motion trajectories that serve as guidance prior. \textcolor{blue(munsell)}{Right: Physics-Guided Score Distillation.} Our Neural Dynamics Model is conditioned on reference trajectories from physics simulation and optimized through Video-SDS with physics regularization losses. This joint optimization refines \textcolor{Purple}{motion} for photorealism while synthesizing \textcolor{blue(munsell)}{appearance}, maintaining physical plausibility through \textcolor{Purple}{physics guidance}. This produces photorealistic \textcolor{Purple}{dynamic} weather effects with continuous particle emission that integrate seamlessly with the static scene.}
\label{fig:overview}
    \vspace{-4mm}
\end{figure*}

\section{Related Work}
\label{sec:related}

\noindent \textbf{Score Distillation for 4D Generation and Editing.} Score Distillation Sampling (SDS)~\cite{poole2022dreamfusion} revolutionized 3D content generation by leveraging 2D diffusion priors to optimize 3D representations. This concept was extended to the temporal domain with Video Score Distillation Sampling (Video-SDS), which uses supervision from text-to-video diffusion models~\cite{singer2023text} to optimize dynamic 3D content. Building on these foundations, recent 4D generation approaches~\cite{ren2023dreamgaussian4d, jiang2025animate3d, zhao2023animate124, ling2024align}, works like Gaussians2Life~\cite{wimmer2024gaussians}, and 4D editing methods~\cite{mou2024instruct, kwon2025efficient} enable text-guided editing of dynamic scenes. However, while these data-driven approaches generate photorealistic appearance, they struggle with physically plausible motion. These methods operate on fixed Gaussian sets, making them unsuitable for weather effects that require continuous particle emission, accumulation, and removal throughout the temporal sequence. 

\smallskip \noindent \textbf{Static 3D Scene Editing.} Recent advances in static neural scene editing have enabled text-driven modifications of 3D representations. Methods such as~\cite{sella2023vox, haque2023instruct, liu2024genn2n, zhuang2023dreameditor} leverage diffusion models to modify rendered content. For Gaussian-based scenes specifically, GaussianEditor~\cite{wang2024gaussianeditor} and GaussCtrl~\cite{wu2024gaussctrl} enable text-guided editing. Environmental editing methods like ClimateNeRF~\cite{li2023climatenerf} enable the addition of static weather effects to neural radiance fields. The key limitation across these approaches is their inability to incorporate temporal dynamics; they either modify existing elements or introduce new static components, but cannot create scene-wide elements that evolve naturally over time.

\smallskip \noindent \textbf{Physics-Based Animation.} Physics-based methods integrate simulation with neural representations, providing strong motion priors through physically accurate dynamics. Methods like PhysGaussian~\cite{xie2024physgaussian}, PAC-NeRF~\cite{li2023pac}, and DreamPhysics~\cite{huang2024dreamphysics} animate existing scene elements through simulation. Gaussian Splashing~\cite{feng2024gaussian} incorporates Position-Based Dynamics for fluid simulation. RainyGS~\cite{dai2025rainygs} synthesizes rain effects through physics-based particle systems, however it can not generalize to additional weather effects. 
While these approaches generate physically-plausible motion, their guidance is derived purely from physics simulation. This motion-only prior is insufficient for photorealism, as these methods cannot synthesize the complex photorealistic appearance for newly introduced dynamic elements, a key limitation our joint optimization framework addresses.

\section{Method}
\label{sec:method}

In this section we introduce our \textit{Physics-Guided Score Distillation} framework for editing 3D Gaussian Splatting scenes scene-wide, multi-particle edits. As illustrated in Fig.~\ref{fig:overview}, our approach consists of two components: First, we use Material Point Method (MPM) simulation to generate physically plausible reference trajectories for dynamic particles (Sec.~\ref{sec:motion_stage}). These trajectories serve as a guidance prior rather than fixed motion. Second, we learn a recurrent neural dynamics model that predicts dynamic Gaussian evolution over time, optimized through Physics-Guided Score Distillation (Sec.~\ref{sec:appearance_stage}), a joint loss combining Video Score Distillation Sampling (Video-SDS) for photorealistic refinement of both motion and appearance with physics regularization for motion plausibility. 
We begin by providing a brief background about the foundational techniques (Sec.~\ref{sec:preliminaries}) that form the basis of our approach. 

\subsection{Preliminaries}
\label{sec:preliminaries}

\noindent \textbf{3D Gaussian Splatting.}
3D Gaussian Splatting (3DGS)~\cite{kerbl20233d} represents scenes using a set of 3D anisotropic Gaussian kernels. Each Gaussian kernel $\mathcal{G}_i$ is defined by its center position $\mathbf{x}_i$, opacity $\sigma_i$, covariance matrix $\mathbf{\Sigma}_i$ which is decomposed into a scaling factor $\mathbf{S}_i \in \mathbb{R}^3$ and a rotation factor $\mathbf{R}_i \in \mathbb{R}^4$, and spherical harmonic coefficients $\mathbf{C}_i$ for view-dependent appearance. The color $\mathbf{C}$ of a pixel is computed by alpha-blending these 3D Gaussians when projected to the image plane:
\begin{equation}
\label{eq:3dgs}
   \mathbf{C} = \sum_{i=1}^N T_i \alpha_i \mathbf{C}_i, \ \rm{with}\ T_i=\prod_{j=1}^{i-1} (1-\alpha_j),
\end{equation}
where $N$ is the set of depth-sorted Gaussian kernels affecting the pixel, and $C_i$ and $\alpha_i$ represents the color and density of this point computed by a 3D Gaussian $G$ with covariance $\mathbf{\Sigma}$ and opacity $\sigma$.

\smallskip
\noindent \textbf{Material Point Method.}
The Material Point Method (MPM)~\cite{stomakhin2013material, jiang2016material} is a numerical simulation technique for continuum dynamics. In MPM, a continuum is represented by a set of particles in a grid-based space. Following PhysGaussian~\cite{xie2024physgaussian}, particles' positions are updated over time:
\begin{equation}
\label{eq:particle_rotation}
   \mathbf{x}_i(t) = \Delta (\mathbf{x}_i, t)
\end{equation}
where $\Delta(\cdot,t)$ updates the particle position. MPM is utilized for its ability to efficiently handle both particle dynamics and physical interactions. 

\smallskip
\noindent \textbf{Video Score Distillation Sampling.}
Score Distillation Sampling (SDS)~\cite{poole2022dreamfusion} distills pre-trained 2D image diffusion models to optimize 3D representations. Video Score Distillation Sampling (Video-SDS)~\cite{singer2023text} extends this to pre-trained text-to-video diffusion model for dynamic 3D generation. Given a camera trajectory $\mathbf{r}(t)$, Video-SDS optimizes the rendered 3D video $V_{\mathbf{r}(t)}$ with predicted noise $\hat{\epsilon}_{\text{V}}$:
\begin{equation}
\label{eq:sdst}
\nabla_{\theta}\mathcal{L}_{\text{Video-SDS}}(\theta) \triangleq \mathbb{E}\left[
    \omega(\mu)
    \left( \hat{\epsilon}_{\text{V}}(V_{\mathbf{r}(t)};\mu,y) - \epsilon \right)
    \frac{\partial V_{\mathbf{r}(t)}}{\partial \theta}
\right], 
\end{equation}
where $\mu$ is noise timestep and $\theta$ are  optimizable parameters.

\subsection{Physics-Based Motion Prior}
\label{sec:motion_stage}

We use MPM simulation to generate reference trajectories that serve as a guidance prior for our optimization. A static scene is first reconstructed from multi-view images using 3DGS to obtain a set of 3D Gaussians $\{\mathcal{G}_i\} = \{\mathbf{x}_i, \sigma_i, \mathbf{S}_i, \mathbf{R}_i, \mathbf{C}_i\}$. We extract a mesh representation of the scene and map the static Gaussians to a particle-based representation suitable for MPM simulation. These static particles act as obstacles in the simulation domain. 

Dynamic particles representing weather effects are then introduced with appropriate physical parameters (emission region, rate, initial velocity, material properties). MPM simulation computes the motion trajectories $\{\mathbf{x}_p^{\text{sim}}(t), \mathbf{v}_p^{\text{sim}}(t)\}$ for these dynamic particles. For computational efficiency, we implement active particle tracking:
\begin{equation}
\label{eq:active_tracking}
a_i^{t+1} = 
\begin{cases}
0, & \text{if } \|\boldsymbol{p}_i^{t+1} - \boldsymbol{p}_i^{t}\| < \delta \text{ or } \boldsymbol{p}_i^{t+1} \notin \Omega \\
a_i^{t}, & \text{otherwise}
\end{cases}
\end{equation}
where $a_i^t$ represents the active state of particle $i$ at time $t$, $\delta$ is a movement threshold, and $\Omega$ is the valid simulation domain. This enables the simulation of thousands of particles across timesteps by efficiently removing particles from active simulation when they come to rest after collision or leave the bounds of interest.

Due to the coarse grid resolution of MPM, we refine particle-scene interactions through mesh-based collision handling. For snow, we project particles onto mesh surfaces and interpolate with nearby surface Gaussians to create natural accumulation. For rain, we implement a 3D wetness grid that tracks moisture with Gaussian smoothing and temporal decay. For sandstorms, we displace particles along surface normals with anisotropic scaling to create thin accumulations. When particles become inactive ($a_i^{t} = 1$ to $a_i^{t+1} = 0$), their final positions and rotations are recorded as collision states. These simulated trajectories provide physically plausible reference motion that guides our neural dynamics model. Additional collision handling details are provided in the supplementary (Sec.~\ref{sec:ch_details}) and simulation parameters in Sec.~\ref{sec:simulation_params}.

\subsection{Physics-Guided Score Distillation}
\label{sec:appearance_stage}

Rather than treating physics simulation as fixed motion, we use it as a guidance prior to optimize a neural dynamics model through Video-SDS. For each dynamic particle from the MPM simulation, we create a corresponding dynamic Gaussian kernel $\mathcal{G}_g(t) = \{\mathbf{x}_g^{\text{init}}(t), \mathbf{R}_g^{\text{init}}, \mathbf{A}_g^{\text{active}}, \mathbf{A}_g^{\text{collided}}, \mathbf{v}_g^{\text{init}}(t)\}$, where $\mathbf{v}_g^{\text{init}}(t)$ is the velocity of each gaussian that is provided along the position $\mathbf{x}_g^{\text{init}}(t)$ directly from the motion simulation. Each Gaussian also maintains state-dependent appearance parameters based on its physical state determined by the physics simulation: $\mathcal{A}_g^{\text{active}}= \{sigma_g^{\text{active}}, \mathbf{S}_g^{\text{active}}, \mathbf{C}_g^{\text{active}}\}$ for actively moving Gaussians and $\mathcal{A}_g^{\text{collided}}$ for Gaussians that have accumulated on static geometry. For notational clarity, we omit the explicit state subscripts and use $\mathcal{A}_g =  \{\sigma_g^{\text{collided}}, \mathbf{S}_g^{\text{collided}}, \mathbf{C}_g^{\text{collided}}\}$ to refer to the appropriate appearance parameters based on the Gaussian's current physical state. Appearance parameters are initialized using a large language model (LLM) that generates appropriate baseline values $\mathcal{A}_g^{\text{init}}$ from text prompts describing the desired weather effect. Additional details on LLM initialization are provided in the supplementary (Sec.~\ref{sec:appearance_init}).

\smallskip
\noindent \textbf{Recurrent Neural Dynamics Model.}
To model the continuous evolution of the dynamic Gaussians, we learn a recurrent neural dynamics model that predicts dynamic Gaussian evolution over time. At each timestep $t$, the model takes as input: (1) the previous rendered state $\{\mathbf{x}_g(t-1), \mathbf{R}_g(t-1), \mathcal{A}_g(t-1)\}$ from the previous timestep, where $\mathbf{x}_g(t-1)$ is the position of Gaussian $g$ at time $t-1$, $\mathbf{R}_g(t-1)$ is the rotation of Gaussian $g$ and $\mathcal{A}_g$ represents the appearance parameters of Gaussian $g$ (opacity, scale, and color), (2) the physics velocity $\mathbf{v}_g^{\text{init}}(t)$ of Gaussian $g$ at time $t$ from simulation, and (3) the current timestep $t$. We implement separate MLPs for each physical state, with each MLP processing Gaussians independently. The positions, velocities, and timestep are processed through learned embedders (Fourier feature encoding for spatial inputs, sinusoidal encoding for time) before being concatenated with the current appearance parameters $\mathcal{A}_g(t-1)$ and fed to the appropriate state-specific MLP. The recurrence is initialized at $t=0$ using physics simulation positions $\mathbf{x}_g^{\text{init}}(0)$ and the LLM-initialized appearance parameters $\mathcal{A}_g^{\text{init}}$. Additional details on the network architecture are provided in the supplementary (Sec.~\ref{sec:imp_details}).

For each Gaussian $g$, the network predicts: (1) velocity correction $\boldsymbol{\Delta}\mathbf{v}_g$ relative to the simulated velocity, (2) angular velocity $\boldsymbol{\omega}_g$ for rotation updates, and (3) appearance deltas $\boldsymbol{\Delta}\mathcal{A}_g$. The refined motion and appearance are computed as:
\begin{equation}
\label{eq:residual_motion}
\begin{aligned}
\mathbf{v}_g(t) &= \mathbf{v}_g^{\text{init}}(t) + \boldsymbol{\Delta}\mathbf{v}_g, \\
\mathbf{x}_g(t) &= \mathbf{x}_g(t-1) + \mathbf{v}_g(t), \\
\mathbf{R}_g(t) &= \mathbf{R}_g(t-1) \circ \Delta \mathbf{R}(\boldsymbol{\omega}_g), \\
\mathcal{A}_g(t) &= \mathcal{A}_g(t-1) + \boldsymbol{\Delta}\mathcal{A}_g,
\end{aligned}
\end{equation}
where $\mathbf{v}_g^{\text{init}}(t)$ serves as a motion prior that is refined by the learned correction, $\circ$ denotes quaternion multiplication and $\Delta \mathbf{R}$ converts angular velocity to a quaternion delta, and appearance parameters evolve recurrently through additive updates. 
The recurrent model, however, introduces drift and error accumulation in the recurrent predictions as the learned trajectories $\mathbf{x}_g(t)$ diverge from the simulation trajectories, and $\mathbf{v}_g^{\text{init}}(t)$ may correspond to a different position in the physics simulation than the actual Gaussian position. The physics regularization losses introduced in the next section are therefore essential to mitigate this drift and maintain physical plausibility throughout optimization.

\smallskip
\noindent \textbf{Physics-Guided Score Distillation Optimization.}
Next, we describe how our recurrent neural dynamics model is optimized through \emph{Physics-Guided Score Distillation}, where the physics prior guides the optimization objective. 

The optimization consists of three objectives. First, the \emph{Video-SDS loss} $\mathcal{L}_{\text{Video-SDS}}$ (Eq.~\ref{eq:sdst}) provides photorealistic guidance from the video diffusion model, ensuring that rendered videos match the target text prompt. Second, \emph{physics guidance losses} regularize the learned motion to remain close to the physics prior through three regularization terms:
\begin{equation}
\label{eq:physics_guidance}
\begin{aligned}
\text{Position: } \mathcal{L}_{\text{xyz}} &= \|\mathbf{x}_g(t) - \mathbf{x}_g^{\text{sim}}(t)\|_{2}, \\
\text{Velocity: } \mathcal{L}_{\text{vel}} &= \|\mathbf{v}_g(t) - \mathbf{v}_g^{\text{sim}}(t)\|_{2}, \\
\text{Rotation: } \mathcal{L}_{\text{rot}} &= 2 \cdot \arccos(|\mathbf{R}_g(t) \cdot \mathbf{R}_g^{\text{init}}|,
\end{aligned}
\end{equation}
where $\mathbf{R}_g(t) \cdot \mathbf{R}_g^{\text{init}}$ is the dot product between two quaternion vectors. The position and velocity regularization losses calculate the $L2$ distance, preventing rendered trajectories from drifting away from the reference simulation trajectories. The rotation regularization loss calculates the angular distance between the two rotations, preventing excessive rotational drift from initial orientations. 

Third, \emph{appearance regularization} mitigates error accumulation inherent to the recurrent formulation by penalizing large appearance deltas:
\begin{equation}
\label{eq:app_reg}
\mathcal{L}_{\text{app}} = \|\boldsymbol{\Delta}\mathcal{A}_g\|_{2} = \|\boldsymbol{\Delta}\sigma_g\|_{2} + \|\boldsymbol{\Delta}\mathbf{S}_g\|_{2} + \|\boldsymbol{\Delta}\mathbf{C}_g\|_{2},
\end{equation}
applied separately for active and collided Gaussian states. The complete objective combines these components with state-specific weights:
\begin{equation}
\label{eq:joint_loss}
\mathcal{L}_{\text{total}} = \mathcal{L}_{\text{Video-SDS}} + \lambda_{\text{xyz}} \mathcal{L}_{\text{xyz}} + \lambda_{\text{vel}} \mathcal{L}_{\text{vel}} + \lambda_{\text{rot}} \mathcal{L}_{\text{rot}} + \lambda_{\text{app}} \mathcal{L}_{\text{app}}.
\end{equation}
Each regularization term has independent weight parameters to provide control over physics-appearance trade-off. Additional details are provided in the supplementary (Sec.~\ref{sec:imp_details}).

These regularization terms are critical for mitigating the drift and error accumulation discussed earlier: higher regularization weights pull the optimization closer to pure physics simulation (physically plausible but potentially lacking photorealism), while lower weights allow the optimization to be dominated by the Video-SDS objective (photorealistic but risking physical implausibility and trajectory drift). To balance this trade-off, we apply SDS-adaptive physics guidance, where all regularization weights are dynamically scaled by the instantaneous Video-SDS loss magnitude $\lambda_i \cdot |\mathcal{L}_{\text{Video-SDS}}|$. This creates an adaptive mechanism where physics guidance is amplified when the diffusion model exhibits high uncertainty (large $|\mathcal{L}_{\text{Video-SDS}}|$) and relaxes when the model is confident (small $|\mathcal{L}_{\text{Video-SDS}}|$), balancing physical plausibility with visual quality.

During training, we render video clips from random camera trajectories, compute $\mathcal{L}_{\text{total}}$, and backpropagate gradients to update network parameters. We employ progressive timestep annealing for the diffusion model to refine details. Through this framework, where physics both conditions network inputs and guides optimization, our model learns to generate physically plausible, photorealistic dynamic weather effects with continuous particle emission.

\begin{figure*}
  \centering
  \includegraphics[width=\textwidth, trim={2.6cm 15.5cm 4.0cm 3.2cm},clip]{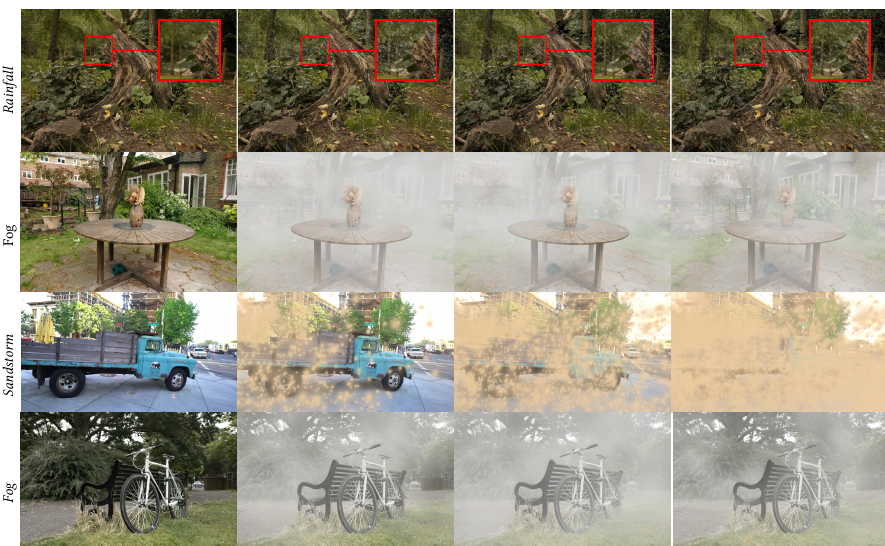}
  \caption{Dynamic weather effects created with our physics-guided SDS framework across different scenes (input frames on left). Each column represents an increasing timestep, demonstrating temporal evolution with continuous particle emission and realistic scene interactions. Our approach enables scene-wide modifications where emitted particles interact naturally with static geometry and evolve coherently over time. The \textit{snowfall} effect is presented in Fig.~\ref{fig:teaser}. As the \textit{rainfall} effect is challenging to visualize, we provide zoomed-in regions to better illustrate the effect. Complete temporal visualizations for all effects are available in the supplementary (Sec.~\ref{sec:interactive}).}
\label{fig:results}
\vspace{-5mm}
\end{figure*}

\section{Experiments}
\label{sec:experiments}

We evaluate our framework for dynamic editing of static 3D Gaussian Splatting scenes through physics-guided score distillation. We present qualitative results across various scenes and effect types in Sec.~\ref{sec:results}, compare against prior work in Sec.~\ref{sec:comparisons}, and provide ablation studies in Sec.~\ref{sec:ablations}. Finally, we discuss limitations in Sec.~\ref{sec:limitations}. 

\smallskip
\noindent \textbf{Scenes and dynamic weather effects.}
We demonstrate our framework on five diverse real-world static scenes: \textit{Garden}, \textit{Bicycle}, and \textit{Stump} from the Mip-NeRF360 dataset~\cite{barron2022mip}, and \textit{Playground} and \textit{Truck} from the Tanks and Temples dataset~\cite{knapitsch2017tanks}. For each scene, we create four types of dynamic weather effects: \textit{snowfall}, \textit{rainfall}, \textit{fog}, and \textit{sandstorm}. This diverse set of effects showcases our approach's ability to handle various dynamic scene modifications with different physical characteristics and temporal behaviors while maintaining physically plausible motion. Additionally, we demonstrate that our framework can generate creative appearance variations through text guidance (\textit{purple snow}, \textit{glittering yellow sand}, and \textit{magic particles}). 

\subsection{Qualitative Results}
\label{sec:results}

\begin{figure}
  \centering
    \includegraphics[width=0.5\textwidth,clip]{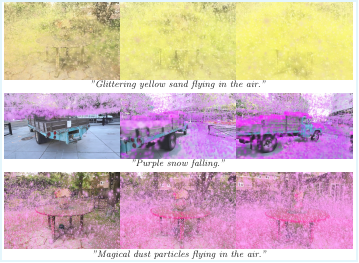}
  \caption{\textbf{Text Prompt Results.} 
   Our physics-guided framework supports creative text prompts beyond standard weather. The method successfully synthesizes the target appearance while maintaining the physically plausible motion learned for the corresponding effect (e.g., sandstorm, snowfall).}
  \vspace{-0.4cm}

\label{fig:results_text}
\end{figure}

Fig.~\ref{fig:teaser} (\textit{snowfall}) and Fig.~\ref{fig:results} (all other dynamic weather effects) demonstrate our method's ability to create visually convincing and temporally consistent dynamic scene modifications across different environments. Our physics-guided score distillation framework enables effects that exhibit physically plausible motion while maintaining photorealistic appearance: snowflakes gently accumulate on surfaces through continuous particle emission, raindrops create wetness patterns, fog diffuses through the scene, and sandstorms envelope objects with physically-driven particle motion. Additionally, Fig.~\ref{fig:results_text} showcases creative appearance variations generated by our framework through text guidance (e.g., \textit{purple snow}) while maintaining physically plausible motion through our physics-guided optimization. Please refer to the supplementary (Sec.~\ref{sec:interactive}) for additional visualizations across different scenes, effects, and viewpoints.

\subsection{Comparison To Baselines}
\label{sec:comparisons}

\begin{figure*}
  \centering
  \includegraphics[width=\textwidth, trim={2.7cm 20.5cm 4.0cm 3.2cm},clip]{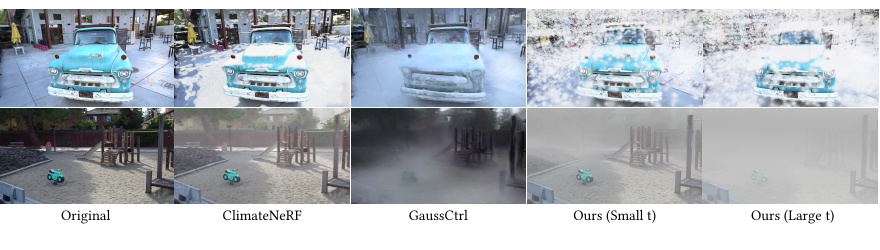}
  \vspace{-1.2cm}
  \caption{\textbf{Qualitative comparison between our dynamic approach and static editing}. 
  We show two different effects across separate rows, comparing the original scene, ClimateNeRF~\cite{li2023climatenerf}, GaussCtrl~\cite{wu2024gaussctrl}, and our method at both early (Small t) and later (Large t) timesteps. 
  }
\label{fig:comparisons}
\vspace{-0.2cm}

\end{figure*}

\begin{figure*}
  \centering
  \includegraphics[width=\textwidth, trim={2.65cm 21.8cm 3.75cm 3.3cm},clip]{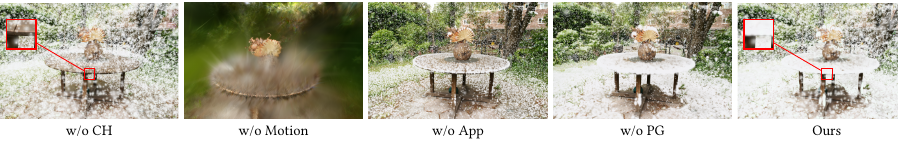}
  \vspace{-1.1cm}
\caption{\textbf{Qualitative Ablation Study}. We compare our full method against four variants. \emph{(w/o CH)} Gaussians float above surfaces without collision handling; \emph{(w/o Motion)} data-driven optimization alone yields incoherent and physically implausible motion; \emph{(w/o App)} physics-only achieves plausible motion but lacks photorealistic appearance; \emph{(w/o PG)} fixed physics motion without joint optimization prevents effective Video-SDS optimization, hindering appearance quality. Our full method achieves both physically plausible motion and photorealistic appearance through physics-guided score distillation. Highlighted regions contrast the successful surface accumulation in our method with the floating artifacts in w/o CH.}
\label{fig:ablations}
\vspace{-4mm}
\end{figure*}

\begin{table}[t]
\centering
\caption{\textbf{Quantitative Evaluation}.  We compare against the static 3D editing methods ClimateNeRF~\cite{li2023climatenerf} and GaussCtrl~\cite{wu2024gaussctrl}, and the 4D editing method Instruct-4DGS~\cite{kwon2025efficient}. Results are shown for different edit categories, with F+S representing results for \textit{fog} and \textit{snow} edits (the only effects supported by ClimateNeRF) and All representing results across all edit types.
}
\resizebox{0.99\linewidth}{!}{\begin{tabular}{llccccc}
\toprule
 &Method  & $\text{CLIP}_{Sim}\uparrow$ & $\text{CLIP}_{Dir} \uparrow$ & $\text{VQAScore}\uparrow$ & $\text{ViCLIP-T}\uparrow$ & $\text{VE-Bench}\uparrow$ \\
\midrule
\multirow{3}{*}{\rotatebox[origin=c]{90}{F+S}} & ClimateNeRF &  $0.23$ & $0.07$ & 0.87 & 0.15 & 0.28 \\
& GaussCtrl &  $0.25$ & $0.08$ & 0.71 & 0.16 & 0.24 \\
& Ours & $\mathbf{0.29}$ & $\mathbf{0.12}$ & $\mathbf{0.92}$ & $\mathbf{0.20}$ & $\mathbf{0.45}$ \\
\midrule 
\multirow{2}{*}{\rotatebox[origin=c]{90}{All}} & GaussCtrl &  $0.24$ & $0.07$ & $0.64$& 0.15 & 0.21 \\
& Ours & $\mathbf{0.28}$ & $\mathbf{0.11}$ & $\mathbf{0.89}$ & $\mathbf{0.19}$ & $\mathbf{0.41}$ \\
\midrule
\multirow{2}{*}{\rotatebox[origin=c]{90}{4D}} & Instruct-4DGS & $0.21$ & $0.07$ & $0.57$ & $0.13$ & $0.19$ \\
& Ours & $\mathbf{0.28}$ & $\mathbf{0.11}$ & $\mathbf{0.89}$ & $\mathbf{0.19}$ & $\mathbf{0.41}$ \\

\bottomrule
\end{tabular}
}
\vspace{-4mm}

\label{tab:comparisons}
\end{table}

\noindent \textbf{Baselines.} 
Our work addresses the challenge of introducing dynamic weather effects to static scenes. We compare against static 3D editing methods: ClimateNeRF~\cite{li2023climatenerf}, which adds static weather effects to NeRFs, and GaussCtrl~\cite{wu2024gaussctrl}, a method for text-guided editing of 3D Gaussian Splatting scenes. We also compare against Instruct-4DGS~\cite{kwon2025efficient}, a recent 4D editing method that applies text-guided edits to dynamic Gaussian scenes via score distillation. Since Instruct-4DGS assumes an existing dynamic scene, we initialize its deformation field to zero and train it jointly, allowing the method to learn motion dynamics from diffusion guidance (additional details in the supplementary, Sec.~\ref{sec:supp_comp}). RainyGS~\cite{dai2025rainygs} addresses rain synthesis through physics-based particle systems with rain-specific simulation and rasterization techniques, however, it is not directly comparable as it does not generalize beyond rain effects and their code is not publicly available.

\smallskip
\noindent \textbf{Evaluation Metrics.} 
We employ metrics that assess visual fidelity and alignment with the intended edit. Our evaluation includes comparisons to static methods; therefore, we consider both image-based and video-based metrics:
(1). 
\emph{CLIP Similarity}~\cite{radford2021learningtransferablevisualmodels}, ($\text{CLIP}_{Sim}$), measures the CLIP similarity between the output rendered images and the target text prompts.
(2). \emph{CLIP Direction Similarity}~\cite{gal2022stylegan} ($\text{CLIP}_{Dir}$), 
measures the cosine distance between the direction of the change from the input and output rendered images and the direction of the change from an input prompt to the edit prompt.
(3). \emph{VQAScore}~\cite{lin2024evaluating}, measures the probability that the answer to the question "Does the image show caption" is "Yes" by inputting the rendered image and
caption into a VQA model.
(4). \emph{VE-Bench Score}~\cite{sun2024bench}, provides a score that evaluates video editing quality by measuring temporal consistency, visual quality, and edit fidelity for dynamic video content through automated metrics designed specifically for assessing video-based edits.
(5). \emph{ViCLIP-T}, as presented in~\cite{yariv2025through}, uses ViCLIP~\cite{wang2024internvidlargescalevideotextdataset}, a video CLIP model that incorporates temporal information when processing videos. ViCLIP-T calculates the cosine similarity between text and video embeddings, measuring how well the output video aligns with the target text prompt, providing a measure of text faithfulness for dynamic content.

\smallskip
\noindent \textbf{Comparisons.}
A qualitative comparison to static 3D editing methods is shown in Fig.~\ref{fig:comparisons}. We show our results at two different timesteps: one at an early stage of the simulation (Small $t$) and one at the end (Large $t$). In the \textit{snowfall} example in the first row, the static methods cannot capture the temporal dynamics of continuous particle emission and accumulation, yielding only surface modifications (ClimateNeRF) or appearance changes (GaussCtrl). In the second row, for the \textit{fog} example, the effect is either unnoticeable (ClimateNeRF) or unrealistic (GaussCtrl). Our physics-guided framework produces effects that exhibit both physically plausible temporal evolution and photorealistic appearance. Additional comparisons across different scenes, effects, and viewpoints are provided in the supplementary (Sec.~\ref{sec:supp_comp}, Sec.~\ref{sec:supp_vis}).

In Tab.~\ref{tab:comparisons} we show a quantitative comparison to to these baselines. For the static methods, we create moving camera videos from their results to enable fair comparison using video-based metrics. ClimateNeRF only performs modifications for the \textit{fog} and \textit{snow} effects, thus we show results over these effects (F+S) and also over all weather effects. As illustrated in the table, our method achieves higher scores across both video and image-based metrics, demonstrating that our approach produces high-quality effects even in static frames, while the temporal dimension, enabled by continuous particle emission and physics-guided optimization, further distinguishes our method. We additionally compare against Instruct-4DGS in Tab.~\ref{tab:comparisons} (4D). Without a physics-based motion prior, Instruct-4DGS produces incoherent results with physically implausible motion, resulting in significantly lower scores across all metrics (qualitative comparison in the supplementary, Fig.~\ref{fig:comparison_4d}).

\subsection{Ablations}
\label{sec:ablations}
We conduct ablation studies to evaluate four variants: (1) without collision handling \emph{(w/o CH)}, removing mesh-based collision handling techniques; (2) without appearance optimization \emph{(w/o App)}, using only physics simulation with LLM-initialized appearance parameters and no Video-SDS optimization; (3) without motion simulation \emph{(w/o Motion)}, using only data-driven optimization by initializing a 4D Gaussian Splatting representation~\cite{wu20244d} and optimizing its deformation field using Video-SDS; and (4) without physics guidance \emph{(w/o PG)}, where motion trajectories from physics simulation are fixed and only appearance parameters are optimized through Video-SDS, eliminating the joint optimization, physics regularization losses, and physics conditioning that enable our approach.

Fig.~\ref{fig:ablations} and Tab.~\ref{tab:ablations} illustrate that each ablated component leads to a significant degradation in quality. The \emph{w/o CH} ablation demonstrates that without proper collision handling, dynamic Gaussians fail to interact convincingly with static scene geometry, with snowflakes floating unnaturally above surfaces instead of accumulating realistically. While the \emph{w/o App} variant achieves physically plausible motion, it produces suboptimal visual appearance due to simplified LLM-initialized parameters. The \emph{w/o Motion} variant produces incoherent and noisy results, producing significantly lower results than other variants. These results demonstrate that video diffusion models alone cannot reliably generate physically plausible motion for complex multi-particle scenarios and are not designed to introduce new Gaussians. The \emph{w/o PG} variant demonstrates that fixed motion prevents effective Video-SDS optimization. Without the joint optimization enabled by our physics-guidance prior, the model cannot learn the motion refinements required for the diffusion model to satisfy the text prompt. Additional ablation results are provided in the supplementary (Sec.~\ref{sec:supp_comp}).

\subsection{Limitations}
\label{sec:limitations}

While our physics-guided score distillation framework achieves high-quality results,
several limitations remain. First, our current approach focuses on adding dynamic elements that interact with static scene geometry but does not enable bi-directional interactions where dynamic effects cause deformation or movement of existing scene elements. Extending our framework to support such interactions represents a promising direction for future work. Second, our framework does not currently update the appearance of static Gaussians to reflect environment changes induced by dynamic effects, such as shadows or ambient lighting variations. The modification of static Gaussian appearance to reflect these environmental effects represents an important direction for future research. Third, while physics simulation provides a strong initialization for optimization, our approach assumes that learned trajectories remain close to the reference simulation. If the optimized trajectories deviate significantly from the physics prior beyond a certain threshold, the guidance velocity signal becomes unreliable. Exploring ways to adapt or recalibrate the guidance
is an interesting direction for future work.

\begin{table}[t]
\centering
\caption{\textbf{Quantitative Ablation Results}. Comparison against variants without collision handling (w/o CH), appearance optimization (w/o App), without motion simulation (w/o Motion) or physical guidance (w/o PG).}
\small
\resizebox{0.99\linewidth}{!}{\begin{tabular}{lccccc}
\toprule
 Method  & $\text{CLIP}_{Sim}\uparrow$ & $\text{CLIP}_{Dir} \uparrow$ & $\text{VQAScore}\uparrow$ & $\text{ViCLIP-T}\uparrow$ & $\text{VE-Bench}\uparrow$\\
\midrule
w/o CH &  0.24 & 0.10 & 0.83 & 0.16  & 0.34 \\
w/o App &  0.25 & 0.10 & 0.82 & 0.16 & 0.37 \\
w/o Motion &  0.18 & 0.03 & 0.35 & 0.08 & 0.13 \\
w/o PG & 0.26 & 0.10 & 0.85 & 0.17 & 0.37 \\
Ours & $\mathbf{0.28}$ & $\mathbf{0.11}$ & $\mathbf{0.89}$ & $\mathbf{0.19}$ &  $\mathbf{0.41}$ \\

\bottomrule
\end{tabular}
}
\vspace{-3mm}
\label{tab:ablations}
\vspace{-3mm}
\end{table}

\section{Conclusion}
\label{sec:conclusion}
We presented \emph{Physics-Guided Score Distillation}, a framework for dynamic editing of 3D Gaussian Splatting scenes. Our key insight is to leverage physics simulation as a powerful motion prior, enabling the joint refinement of both motion and appearance within a unified Video-SDS optimization. Our ablations confirm this joint refinement is essential for achieving coherent, photorealistic results, as simpler alternatives fail. This work bridges physics-based simulation with data-driven diffusion models, opening a new path for realistic, dynamic 3D content creation.

\paragraph{Acknowledgements.} This research was supported by The Israel Science Foundation (grant No. 2416/25).

{
    \small
    \bibliographystyle{ieeenat_fullname}
    \bibliography{main}

@String(TOG= {ACM Trans. Graph.})

@String(TOG   = {ACM TOG})

@article{kerbl20233d,
  title={3d gaussian splatting for real-time radiance field rendering.},
  author={Kerbl, Bernhard and Kopanas, Georgios and Leimk{\"u}hler, Thomas and Drettakis, George},
  journal={ACM Trans. Graph.},
  volume={42},
  number={4},
  pages={139--1},
  year={2023}
}

@inproceedings{li2023climatenerf,
  title={Climatenerf: Extreme weather synthesis in neural radiance field},
  author={Li, Yuan and Lin, Zhi-Hao and Forsyth, David and Huang, Jia-Bin and Wang, Shenlong},
  booktitle={Proceedings of the ieee/cvf international conference on computer vision},
  pages={3227--3238},
  year={2023}
}

@inproceedings{xie2024physgaussian,
  title={Physgaussian: Physics-integrated 3d gaussians for generative dynamics},
  author={Xie, Tianyi and Zong, Zeshun and Qiu, Yuxing and Li, Xuan and Feng, Yutao and Yang, Yin and Jiang, Chenfanfu},
  booktitle={Proceedings of the IEEE/CVF Conference on Computer Vision and Pattern Recognition},
  pages={4389--4398},
  year={2024}
}

@article{li2023pac,
  title={Pac-nerf: Physics augmented continuum neural radiance fields for geometry-agnostic system identification},
  author={Li, Xuan and Qiao, Yi-Ling and Chen, Peter Yichen and Jatavallabhula, Krishna Murthy and Lin, Ming and Jiang, Chenfanfu and Gan, Chuang},
  journal={arXiv preprint arXiv:2303.05512},
  year={2023}
}

@article{qiu2024feature,
  title={Feature splatting: Language-driven physics-based scene synthesis and editing},
  author={Qiu, Ri-Zhao and Yang, Ge and Zeng, Weijia and Wang, Xiaolong},
  journal={arXiv preprint arXiv:2404.01223},
  year={2024}
}

@article{huang2024dreamphysics,
  title={DreamPhysics: Learning physics-based 3D dynamics with video diffusion priors},
  author={Huang, Tianyu and Zhang, Haoze and Zeng, Yihan and Zhang, Zhilu and Li, Hui and Zuo, Wangmeng and Lau, Rynson WH},
  journal={arXiv preprint arXiv:2406.01476},
  year={2024}
}

@inproceedings{zhang2024physdreamer,
  title={Physdreamer: Physics-based interaction with 3d objects via video generation},
  author={Zhang, Tianyuan and Yu, Hong-Xing and Wu, Rundi and Feng, Brandon Y and Zheng, Changxi and Snavely, Noah and Wu, Jiajun and Freeman, William T},
  booktitle={European Conference on Computer Vision},
  pages={388--406},
  year={2024},
  organization={Springer}
}

@article{ren2023dreamgaussian4d,
  title={Dreamgaussian4d: Generative 4d gaussian splatting},
  author={Ren, Jiawei and Pan, Liang and Tang, Jiaxiang and Zhang, Chi and Cao, Ang and Zeng, Gang and Liu, Ziwei},
  journal={arXiv preprint arXiv:2312.17142},
  year={2023}
}

@article{zhao2023animate124,
  title={Animate124: Animating one image to 4d dynamic scene},
  author={Zhao, Yuyang and Yan, Zhiwen and Xie, Enze and Hong, Lanqing and Li, Zhenguo and Lee, Gim Hee},
  journal={arXiv preprint arXiv:2311.14603},
  year={2023}
}

@article{jiang2025animate3d,
  title={Animate3d: Animating any 3d model with multi-view video diffusion},
  author={Jiang, Yanqin and Yu, Chaohui and Cao, Chenjie and Wang, Fan and Hu, Weiming and Gao, Jin},
  journal={Advances in Neural Information Processing Systems},
  volume={37},
  pages={125879--125906},
  year={2025}
}

@inproceedings{ling2024align,
  title={Align your gaussians: Text-to-4d with dynamic 3d gaussians and composed diffusion models},
  author={Ling, Huan and Kim, Seung Wook and Torralba, Antonio and Fidler, Sanja and Kreis, Karsten},
  booktitle={Proceedings of the IEEE/CVF conference on computer vision and pattern recognition},
  pages={8576--8588},
  year={2024}
}

@inproceedings{barron2022mip,
  title={Mip-nerf 360: Unbounded anti-aliased neural radiance fields},
  author={Barron, Jonathan T and Mildenhall, Ben and Verbin, Dor and Srinivasan, Pratul P and Hedman, Peter},
  booktitle={Proceedings of the IEEE/CVF conference on computer vision and pattern recognition},
  pages={5470--5479},
  year={2022}
}

@article{knapitsch2017tanks,
  title={Tanks and temples: Benchmarking large-scale scene reconstruction},
  author={Knapitsch, Arno and Park, Jaesik and Zhou, Qian-Yi and Koltun, Vladlen},
  journal={ACM Transactions on Graphics (ToG)},
  volume={36},
  number={4},
  pages={1--13},
  year={2017},
  publisher={ACM New York, NY, USA}
}

@incollection{jiang2016material,
  title={The material point method for simulating continuum materials},
  author={Jiang, Chenfanfu and Schroeder, Craig and Teran, Joseph and Stomakhin, Alexey and Selle, Andrew},
  booktitle={Acm siggraph 2016 courses},
  pages={1--52},
  year={2016}
}

@inproceedings{huang20242d,
  title={2d gaussian splatting for geometrically accurate radiance fields},
  author={Huang, Binbin and Yu, Zehao and Chen, Anpei and Geiger, Andreas and Gao, Shenghua},
  booktitle={ACM SIGGRAPH 2024 conference papers},
  pages={1--11},
  year={2024}
}

@article{yariv2025through,
  title={Through-The-Mask: Mask-based Motion Trajectories for Image-to-Video Generation},
  author={Yariv, Guy and Kirstain, Yuval and Zohar, Amit and Sheynin, Shelly and Taigman, Yaniv and Adi, Yossi and Benaim, Sagie and Polyak, Adam},
  journal={arXiv preprint arXiv:2501.03059},
  year={2025}
}

@misc{radford2021learningtransferablevisualmodels,
      title={Learning Transferable Visual Models From Natural Language Supervision}, 
      author={Alec Radford and Jong Wook Kim and Chris Hallacy and Aditya Ramesh and Gabriel Goh and Sandhini Agarwal and Girish Sastry and Amanda Askell and Pamela Mishkin and Jack Clark and Gretchen Krueger and Ilya Sutskever},
      year={2021},
      eprint={2103.00020},
      archivePrefix={arXiv},
      primaryClass={cs.CV},
      url={https://arxiv.org/abs/2103.00020}, 
}

@misc{wang2024internvidlargescalevideotextdataset,
      title={InternVid: A Large-scale Video-Text Dataset for Multimodal Understanding and Generation}, 
      author={Yi Wang and Yinan He and Yizhuo Li and Kunchang Li and Jiashuo Yu and Xin Ma and Xinhao Li and Guo Chen and Xinyuan Chen and Yaohui Wang and Conghui He and Ping Luo and Ziwei Liu and Yali Wang and Limin Wang and Yu Qiao},
      year={2024},
      eprint={2307.06942},
      archivePrefix={arXiv},
      primaryClass={cs.CV},
      url={https://arxiv.org/abs/2307.06942}, 
}

@article{feng2024gaussian,
  title={Gaussian Splashing: Unified Particles for Versatile Motion Synthesis and Rendering},
  author={Feng, Yutao and Feng, Xiang and Shang, Yintong and Jiang, Ying and Yu, Chang and Zong, Zeshun and Shao, Tianjia and Wu, Hongzhi and Zhou, Kun and Jiang, Chenfanfu and others},
  journal={arXiv preprint arXiv:2401.15318},
  year={2024}
}

@article{stomakhin2013material,
  title={A material point method for snow simulation},
  author={Stomakhin, Alexey and Schroeder, Craig and Chai, Lawrence and Teran, Joseph and Selle, Andrew},
  journal={ACM Transactions on Graphics (TOG)},
  volume={32},
  number={4},
  pages={1--10},
  year={2013},
  publisher={ACM New York, NY, USA}
}

@inproceedings{poole2022dreamfusion,
  title={DreamFusion: Text-to-3d using 2d diffusion},
  author={Poole, Ben and Jain, Ajay and Barron, Jonathan T and Mildenhall, Ben},
  booktitle={The Eleventh International Conference on Learning Representations},
  year={2022}
}

@inproceedings{singer2023text,
  title={Text-to-4d dynamic scene generation},
  author={Singer, Uriel and Sheynin, Shelly and Polyak, Adam and Ashual, Oron and Makarov, Iurii and Kokkinos, Filippos and Goyal, Naman and Vedaldi, Andrea and Parikh, Devi and Johnson, Justin and others},
  booktitle={Proceedings of the 40th International Conference on Machine Learning},
  pages={31915--31929},
  year={2023},
  organization={PMLR}
}

@article{wimmer2024gaussians,
  title={Gaussians-to-Life: Text-Driven Animation of 3D Gaussian Splatting Scenes},
  author={Wimmer, Thomas and Oechsle, Michael and Niemeyer, Michael and Tombari, Federico},
  journal={arXiv preprint arXiv:2411.19233},
  year={2024}
}

@article{hu2019taichi,
  title={Taichi: a language for high-performance computation on spatially sparse data structures},
  author={Hu, Yuanming and Li, Tzu-Mao and Anderson, Luke and Ragan-Kelley, Jonathan and Durand, Fr{\'e}do},
  journal={ACM Transactions on Graphics (TOG)},
  volume={38},
  number={6},
  pages={1--16},
  year={2019},
  publisher={ACM New York, NY, USA}
}

@inproceedings{wu2024gaussctrl,
  title={Gaussctrl: Multi-view consistent text-driven 3d gaussian splatting editing},
  author={Wu, Jing and Bian, Jia-Wang and Li, Xinghui and Wang, Guangrun and Reid, Ian and Torr, Philip and Prisacariu, Victor Adrian},
  booktitle={European conference on computer vision},
  pages={55--71},
  year={2024},
  organization={Springer}
}

@inproceedings{lin2024evaluating,
  title={Evaluating text-to-visual generation with image-to-text generation},
  author={Lin, Zhiqiu and Pathak, Deepak and Li, Baiqi and Li, Jiayao and Xia, Xide and Neubig, Graham and Zhang, Pengchuan and Ramanan, Deva},
  booktitle={European Conference on Computer Vision},
  pages={366--384},
  year={2024},
  organization={Springer}
}

@article{gal2022stylegan,
  title={Stylegan-nada: Clip-guided domain adaptation of image generators},
  author={Gal, Rinon and Patashnik, Or and Maron, Haggai and Bermano, Amit H and Chechik, Gal and Cohen-Or, Daniel},
  journal={ACM Transactions on Graphics (TOG)},
  volume={41},
  number={4},
  pages={1--13},
  year={2022},
  publisher={ACM New York, NY, USA}
}

@Manual{blender,
   title = {Blender - a 3D modelling and rendering package},
   author = {Blender Online Community},
   organization = {Blender Foundation},
   address = {Stichting Blender Foundation, Amsterdam},
   year = {2018},
   url = {http://www.blender.org},
 }

@article{Zhou2018,
    author    = {Qian-Yi Zhou and Jaesik Park and Vladlen Koltun},
    title     = {{Open3D}: {A} Modern Library for {3D} Data Processing},
    journal   = {arXiv:1801.09847},
    year      = {2018},
}

@inproceedings{dai2025rainygs,
  title={Rainygs: Efficient rain synthesis with physically-based gaussian splatting},
  author={Dai, Qiyu and Ni, Xingyu and Shen, Qianfan and Chen, Wenzheng and Chen, Baoquan and Chu, Mengyu},
  booktitle={Proceedings of the Computer Vision and Pattern Recognition Conference},
  pages={16153--16162},
  year={2025}
}

@inproceedings{haque2023instruct,
  title={Instruct-nerf2nerf: Editing 3d scenes with instructions},
  author={Haque, Ayaan and Tancik, Matthew and Efros, Alexei A and Holynski, Aleksander and Kanazawa, Angjoo},
  booktitle={Proceedings of the IEEE/CVF International Conference on Computer Vision},
  pages={19740--19750},
  year={2023}
}

@inproceedings{zhuang2023dreameditor,
  title={Dreameditor: Text-driven 3d scene editing with neural fields},
  author={Zhuang, Jingyu and Wang, Chen and Lin, Liang and Liu, Lingjie and Li, Guanbin},
  booktitle={SIGGRAPH Asia 2023 conference papers},
  pages={1--10},
  year={2023}
}

@inproceedings{sella2023vox,
  title={Vox-e: Text-guided voxel editing of 3d objects},
  author={Sella, Etai and Fiebelman, Gal and Hedman, Peter and Averbuch-Elor, Hadar},
  booktitle={Proceedings of the IEEE/CVF international conference on computer vision},
  pages={430--440},
  year={2023}
}

@inproceedings{chen2024gaussianeditor,
  title={Gaussianeditor: Swift and controllable 3d editing with gaussian splatting},
  author={Chen, Yiwen and Chen, Zilong and Zhang, Chi and Wang, Feng and Yang, Xiaofeng and Wang, Yikai and Cai, Zhongang and Yang, Lei and Liu, Huaping and Lin, Guosheng},
  booktitle={Proceedings of the IEEE/CVF conference on computer vision and pattern recognition},
  pages={21476--21485},
  year={2024}
}

@inproceedings{wang2024gaussianeditor,
  title={Gaussianeditor: Editing 3d gaussians delicately with text instructions},
  author={Wang, Junjie and Fang, Jiemin and Zhang, Xiaopeng and Xie, Lingxi and Tian, Qi},
  booktitle={Proceedings of the IEEE/CVF conference on computer vision and pattern recognition},
  pages={20902--20911},
  year={2024}
}

@inproceedings{liu2024genn2n,
  title={Genn2n: Generative nerf2nerf translation},
  author={Liu, Xiangyue and Xue, Han and Luo, Kunming and Tan, Ping and Yi, Li},
  booktitle={Proceedings of the IEEE/CVF Conference on Computer Vision and Pattern Recognition},
  pages={5105--5114},
  year={2024}
}

@article{sun2024bench,
  title={VE-Bench: Subjective-Aligned Benchmark Suite for Text-Driven Video Editing Quality Assessment},
  author={Sun, Shangkun and Liang, Xiaoyu and Fan, Songlin and Gao, Wenxu and Gao, Wei},
  journal={arXiv preprint arXiv:2408.11481},
  year={2024}
}

@inproceedings{wu20244d,
  title={4d gaussian splatting for real-time dynamic scene rendering},
  author={Wu, Guanjun and Yi, Taoran and Fang, Jiemin and Xie, Lingxi and Zhang, Xiaopeng and Wei, Wei and Liu, Wenyu and Tian, Qi and Wang, Xinggang},
  booktitle={Proceedings of the IEEE/CVF conference on computer vision and pattern recognition},
  pages={20310--20320},
  year={2024}
}

@article{yang2024cogvideox,
  title={Cogvideox: Text-to-video diffusion models with an expert transformer},
  author={Yang, Zhuoyi and Teng, Jiayan and Zheng, Wendi and Ding, Ming and Huang, Shiyu and Xu, Jiazheng and Yang, Yuanming and Hong, Wenyi and Zhang, Xiaohan and Feng, Guanyu and others},
  journal={arXiv preprint arXiv:2408.06072},
  year={2024}
}

@inproceedings{kwon2025efficient,
  title={Efficient dynamic scene editing via 4d gaussian-based static-dynamic separation},
  author={Kwon, Joohyun and Cho, Hanbyel and Kim, Junmo},
  booktitle={Proceedings of the IEEE/CVF Conference on Computer Vision and Pattern Recognition},
  pages={26855--26865},
  year={2025}
}

@inproceedings{mou2024instruct,
  title={Instruct 4d-to-4d: Editing 4d scenes as pseudo-3d scenes using 2d diffusion},
  author={Mou, Linzhan and Chen, Jun-Kun and Wang, Yu-Xiong},
  booktitle={Proceedings of the IEEE/CVF Conference on Computer Vision and Pattern Recognition},
  pages={20176--20185},
  year={2024}
}

@article{katzir2023noise,
  title={Noise-free score distillation},
  author={Katzir, Oren and Patashnik, Or and Cohen-Or, Daniel and Lischinski, Dani},
  journal={arXiv preprint arXiv:2310.17590},
  year={2023}
}

@article{mcallister2024rethinking,
  title={Rethinking score distillation as a bridge between image distributions},
  author={McAllister, David and Ge, Songwei and Huang, Jia-Bin and Jacobs, David W and Efros, Alexei A and Holynski, Aleksander and Kanazawa, Angjoo},
  journal={Advances in Neural Information Processing Systems},
  volume={37},
  pages={33779--33804},
  year={2024}
}
}

\clearpage
\maketitlesupplementary

\section{Interactive Visualizations}
\label{sec:interactive}

We refer readers to the interactive visualizations at \href{https://galfiebelman.github.io/let-it-snow/supp/index.html}{https://galfiebelman.github.io/let-it-snow/supp/index.html} for full temporal sequences of dynamic weather effects, comparisons to 4D editing baselines, and comprehensive ablation studies.

\section{Additional Ablations and Comparisons}
\label{sec:supp_comp}

\smallskip
\noindent \textbf{4D Editing Baseline Comparison.}
We compare our method against Instruct-4DGS~\cite{kwon2025efficient}, a recent approach for scene-wide, text-guided editing of 4D Gaussian Splatting scenes. Instruct-4DGS operates in three stages: first generating consistent 2D target images by editing multiview renderings using an image-conditioned diffusion model, then optimizing the canonical 3D Gaussian Splatting representation to match these 2D priors through reconstruction loss, and finally refining the canonical representation using diffusion guidance while keeping the deformation field fixed. The method is designed for editing existing dynamic scenes captured from videos, where it modifies the canonical representation while freezing the pre-trained deformation field to preserve the original motion. 

However, Instruct-4DGS is not directly suited for our task of introducing new dynamic weather effects to static scenes, as it assumes an existing dynamic scene and edits only the canonical representation while preserving the original deformation field. To enable comparison, we initialize the deformation field to zero and train it jointly during the refinement stage, allowing the method to learn motion dynamics from diffusion guidance. We use the code provided by the authors (\href{https://github.com/juhyeon-kwon/Instruct-4DGS}{https://github.com/juhyeon-kwon/Instruct-4DGS}) with the default parameters, extending the optimization to 3,000 iterations for the first stage and 5,000 iterations for the second stage to ensure convergence.

\begin{figure*}
  \centering
  \includegraphics[width=\textwidth, trim={3.41cm 18.5cm 4.75cm 3.2cm},clip]{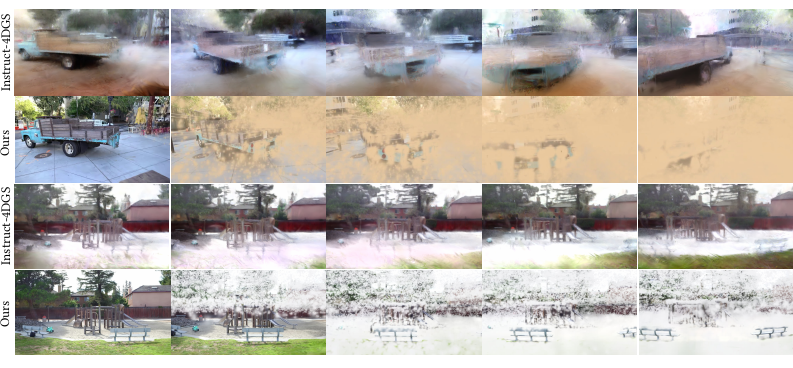}
  \vspace{-0.8cm}
\caption{\textbf{Qualitative Comparison to 4D Editing Baseline}. We compare our method against Instruct-4DGS~\cite{kwon2025efficient} across two dynamic weather effects over increasing timesteps. Top two rows show the \emph{sandstorm} effect, bottom two rows show the \emph{snowfall} effect. Instruct-4DGS produces noisy and incoherent results with physically implausible motion patterns, as the method lacks physics-based guidance to learn realistic multi-particle dynamics. Our physics-guided score distillation framework generates photorealistic weather effects with both plausible motion and photorealistic appearance.}
\label{fig:comparison_4d}
\end{figure*}

As shown in Fig.~\ref{fig:comparison_4d} (quantitative results in Tab.~\ref{tab:comparisons} of the main paper), Instruct-4DGS produces incoherent results with physically implausible motion patterns. Without a physics-based motion prior to guide the optimization, the method struggles to learn realistic dynamics for complex multi-particle weather phenomena, resulting in unnatural motion and visual artifacts.

\smallskip
\noindent \textbf{SDS-Adaptive Weight Ablations.}
We ablate our SDS-adaptive weighting mechanism by replacing all adaptive regularization weights with constant values across both physics guidance losses ($\mathcal{L}_{\text{xyz}}$, $\mathcal{L}_{\text{vel}}$, $\mathcal{L}_{\text{rot}}$) and appearance regularization ($\mathcal{L}_{\text{app}}$). We evaluate four fixed weight values: $\lambda{=}1$, $\lambda{=}100$, $\lambda{=}10^4$, and $\lambda{=}10^6$, comparing against our full method with SDS-adaptive scaling.

\begin{figure*}
  \centering
  \includegraphics[width=\textwidth, trim={2.65cm 21.8cm 3.75cm 3.3cm},clip]{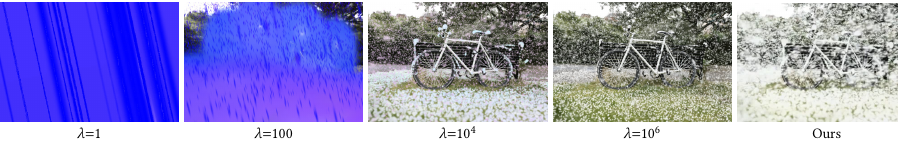}
  \vspace{-1.1cm}
\caption{\textbf{SDS-Adaptive Weight Ablations}. Comparison of constant regularization weights versus our SDS-adaptive approach. $\lambda{=}1$ produces severe visual artifacts and noise due to insufficient regularization; $\lambda{=}100$ shows reduced but still significant noise and color distortions; $\lambda{=}10^4$ exhibits unnatural color artifacts with pinkish tones; $\lambda{=}10^6$ over-regularizes the optimization, preventing refinement and producing results similar to the w/o App ablation (Sec.~\ref{sec:ablations}), where appearance remains close to LLM-initialized parameters. These results demonstrate that fixed regularization weights cannot adapt to the varying and noisy Video-SDS loss throughout optimization. Our SDS-adaptive approach dynamically scales regularization based on instantaneous SDS loss magnitude, automatically balancing physics guidance and appearance stability.}
\label{fig:ablations_adaptive}
\end{figure*}

As shown in Fig.~\ref{fig:ablations_adaptive} and Tab.~\ref{tab:ablations_adaptive}, low constant weights ($1$, $100$) result in noisy and unstable optimization, with weight $1$ producing severe visual artifacts and weight $100$ showing reduced but still significant noise. At moderate weight ($10^4$), we observe color artifacts with unnatural pinkish tones, indicating that constant regularization fails to properly balance physics guidance with photorealistic optimization. Conversely, excessively high constant weight ($10^6$) over-regularizes the optimization, producing results similar to the \emph{w/o App} ablation (Sec.~\ref{sec:ablations}) where appearance remains close to LLM-initialized parameters with minimal refinement. These results demonstrate that fixed regularization weights cannot adequately balance the competing objectives throughout optimization. Because the Video-SDS loss exhibits varying magnitude during training, regularization weights must adapt dynamically to maintain this balance. Our SDS-adaptive approach scales all regularization weights by the instantaneous Video-SDS loss magnitude, automatically adjusting the strength of physics guidance and appearance stability as the diffusion model's uncertainty changes throughout optimization.

\smallskip
\noindent \textbf{Regularization Loss Ablations.}
We conduct ablation studies on each regularization loss component to validate their individual contributions to our physics-guided optimization. We evaluate four variants: (1) without position regularization \emph{(w/o $\mathcal{L}_{\text{xyz}}$)}, (2) without velocity regularization \emph{(w/o $\mathcal{L}_{\text{vel}}$)}, (3) without rotation regularization \emph{(w/o $\mathcal{L}_{\text{rot}}$)}, and (4) without appearance regularization \emph{(w/o $\mathcal{L}_{\text{app}}$)}. 

\begin{figure*}
  \centering
  \includegraphics[width=\textwidth, trim={2.65cm 21.8cm 3.75cm 3.3cm},clip]{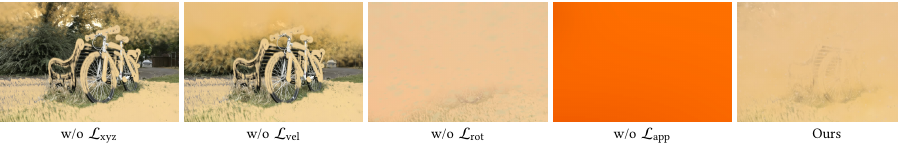}
  \vspace{-1.1cm}
\caption{\textbf{Regularization Loss Ablations}. Comparison of our full method against variants removing individual regularization losses. \emph{w/o $\mathcal{L}_{\text{xyz}}$} and \emph{w/o $\mathcal{L}_{\text{vel}}$} cause trajectory drift from the physics prior, as the sandstorm passes above the bicycle and the bench; \emph{w/o $\mathcal{L}_{\text{rot}}$} shows minimal impact with results similar to our full method, indicating small rotational corrections; \emph{w/o $\mathcal{L}_{\text{app}}$} leads to severe error accumulation with noisy appearance. These results demonstrate that physics regularization is essential for maintaining plausible motion during Video-SDS optimization, while appearance regularization is critical for preventing error accumulation in our recurrent neural dynamics model.}
\label{fig:ablations_reg}
\end{figure*}

As shown in Fig.~\ref{fig:ablations_reg} and Tab.~\ref{tab:ablations_reg}, removing position or velocity regularization causes significant trajectory drift from the physics prior, as the learned motion deviates substantially from the simulated reference trajectories without constraints to maintain proximity. The rotation regularization ablation shows minimal impact, producing results similar to our full method, indicating that rotational corrections remain small throughout optimization. Without appearance regularization, we observe severe error accumulation in the recurrent predictions, resulting in completely noisy and incoherent results.

\begin{figure*}
  \centering
  \includegraphics[width=\textwidth, trim={2.65cm 21.8cm 3.75cm 3.3cm},clip]{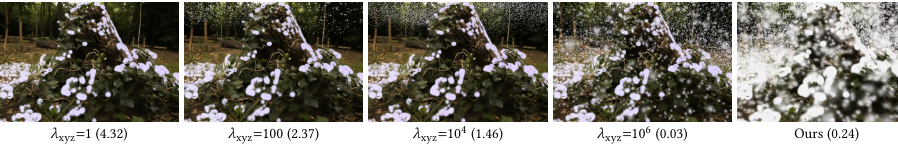}
  \vspace{-1.1cm}
\caption{\textbf{Trajectory Drift Ablations}. Comparison of constant regularization weights for $\mathcal{L}_{\text{xyz}}$ demonstrating threshold behavior. Numbers in parentheses show $\text{Avg. Drift}$ (mean Euclidean distance between rendered and simulated positions). At low weights ($\lambda_{\text{xyz}}{=}1$, $\lambda_{\text{xyz}}{=}100$, $\lambda_{\text{xyz}}{=}10^4$), active snow Gaussians drift significantly from simulation trajectories and fail to reach the ground, resulting in minimal or no visible falling snow (though accumulated snow remains visible as its positions are fixed at collision states). At extremely high weight ($\lambda_{\text{xyz}}{=}10^6$), drift is minimal but optimization is over-constrained, preventing photorealistic refinement. Our SDS-adaptive approach maintains low drift while enabling effective optimization.}
\label{fig:ablations_drift}
\end{figure*}

\smallskip
\noindent \textbf{Trajectory Drift Ablations.}
As discussed in our limitations (Sec.~\ref{sec:limitations}), our physics guidance assumes that learned trajectories remain reasonably close to the simulation reference, thus when trajectories deviate significantly beyond a certain threshold, the guidance signal becomes unreliable. To investigate this threshold behavior, we replace our SDS-adaptive physics guidance with constant regularization weights for position loss $\mathcal{L}_{\text{xyz}}$. We evaluate four fixed weight values: $\lambda_{\text{xyz}}{=}1$, $\lambda_{\text{xyz}}{=}100$, $\lambda_{\text{xyz}}{=}10^4$, and $\lambda_{\text{xyz}}{=}10^6$, comparing against our full method with adaptive weighting. To quantify trajectory drift, we measure the average distance between rendered and simulated positions over all timesteps (reported as "Avg. Drift" in Tab.~\ref{tab:ablations_drift}):
\begin{equation}
\label{eq:drift_eq}
\text{Avg. Drift} =\frac{1}{T} \sum_{t=1}^{T} \frac{1}{N_{t}} \sum_{i=1}^{N_{t}} \| \mathbf{x}_{i}^{\text{rendered}} - \mathbf{x}_{i}^{\text{sim}} \|_2
\end{equation}

where $T$ is the total timesteps and $N_{t}$ is the number of moving Gaussians at timestep $t$.
As shown in Fig.~\ref{fig:ablations_drift} and Tab.~\ref{tab:ablations_drift}, low regularization weights ($1$, $100$, $10^4$) allow trajectories to deviate significantly from the simulation prior, causing high drift values that degrade physics guidance quality. Conversely, excessively high regularization weight ($10^6$) over-constrains the optimization, such that drift becomes minimal but motion remains rigidly fixed to simulation trajectories, preventing the refinements necessary for photorealism (similar to the w/o PG ablation (Sec.~\ref{sec:ablations}). This demonstrates the existence of a critical threshold: insufficient regularization causes unreliable guidance due to excessive drift, while excessive regularization prevents effective optimization. Our SDS-adaptive approach aims to maintain trajectories within this viable range, though developing methods to dynamically recalibrate guidance as trajectories diverge remains an important direction for future work.

\begin{table}[t]
\centering
\caption{\textbf{SDS-Adaptive Weight Ablations}. Quantitative comparison of constant regularization weights versus our SDS-adaptive approach. Low weights ($\lambda{=}1$, $\lambda{=}100$) produce noisy and unstable results, moderate weight ($\lambda{=}10^4$) shows color artifacts, and very high weight ($\lambda{=}10^6$) over-constrains optimization. Our SDS-adaptive approach outperforms all fixed weights by dynamically adjusting to Video-SDS loss magnitude.}
\small
\resizebox{0.99\linewidth}{!}{\begin{tabular}{lccccc}
\toprule
 Method  & $\text{CLIP}_{Sim}\uparrow$ & $\text{CLIP}_{Dir} \uparrow$ & $\text{VQAScore}\uparrow$ & $\text{ViCLIP-T}\uparrow$ & $\text{VE-Bench}\uparrow$\\
\midrule
$\lambda{=}1$ & 0.14 & 0.02 & 0.25 & 0.07  & 0.08 \\
$\lambda{=}100$ &  0.21 & 0.05 & 0.43 & 0.09 & 0.18 \\
$\lambda{=}10^4$ &  0.26 & 0.10 & 0.84 & 0.16 & 0.39 \\
$\lambda{=}10^6$ & 0.25 & 0.10 & 0.83 & 0.16 & 0.36 \\
Ours & $\mathbf{0.28}$ & $\mathbf{0.11}$ & $\mathbf{0.89}$ & $\mathbf{0.19}$ &  $\mathbf{0.41}$ \\

\bottomrule
\end{tabular}
}
\label{tab:ablations_adaptive}
\end{table}

\begin{table}[t]
\centering
\caption{\textbf{Regularization Loss Ablations}. Quantitative comparison of variants removing individual regularization losses: without position regularization \emph{(w/o $\mathcal{L}_{\text{xyz}}$)}, velocity regularization \emph{(w/o $\mathcal{L}_{\text{vel}}$)}, rotation regularization \emph{(w/o $\mathcal{L}_{\text{rot}}$)}, or appearance regularization \emph{(w/o $\mathcal{L}_{\text{app}}$)}. Removing position or velocity regularization causes trajectory drift, while removing appearance regularization leads to severe error accumulation and incoherent results.}

\small
\resizebox{0.99\linewidth}{!}{\begin{tabular}{lccccc}
\toprule
 Method  & $\text{CLIP}_{Sim}\uparrow$ & $\text{CLIP}_{Dir} \uparrow$ & $\text{VQAScore}\uparrow$ & $\text{ViCLIP-T}\uparrow$ & $\text{VE-Bench}\uparrow$\\
\midrule
w/o $\mathcal{L}_{\text{xyz}}$ & 0.26 & 0.09 & 0.83 & 0.16  & 0.33 \\
w/o $\mathcal{L}_{\text{vel}}$ &  0.26 & 0.10 & 0.84 & 0.16 & 0.36 \\
w/o $\mathcal{L}_{\text{rot}}$ & $\mathbf{0.28}$ & $\mathbf{0.11}$ & 0.87 & $\mathbf{0.19}$ & 0.40 \\
w/o $\mathcal{L}_{\text{app}}$ & 0.15 & 0.01 & 0.28 & 0.05 & 0.09 \\
Ours & $\mathbf{0.28}$ & $\mathbf{0.11}$ & $\mathbf{0.89}$ & $\mathbf{0.19}$ &  $\mathbf{0.41}$ \\

\bottomrule
\end{tabular}
}
\label{tab:ablations_reg}
\end{table}

\begin{table}[t]
\centering
\caption{\textbf{Trajectory Drift Ablations}. Investigation of the trajectory drift threshold using constant regularization weights for $\mathcal{L}_{\text{xyz}}$. $\text{Avg. Drift}$ measures average distance between rendered and simulated positions. Results demonstrate a critical threshold: insufficient regularization causes excessive drift that degrades guidance quality, while excessive regularization prevents meaningful optimization.}
\small
\resizebox{0.99\linewidth}{!}{\begin{tabular}{lcccccc}
\toprule
 Method  & \text{Avg. Drift}& $\text{CLIP}_{Sim}\uparrow$ & $\text{CLIP}_{Dir} \uparrow$ & $\text{VQAScore}\uparrow$ & $\text{ViCLIP-T}\uparrow$ & $\text{VE-Bench}\uparrow$\\
\midrule
$\lambda_{\text{xyz}}{=}1$ & 4.32 & 0.25 & 0.09 & 0.80 & 0.15  & 0.32 \\
$\lambda_{\text{xyz}}{=}100$ & 2.37 & 0.26 & 0.09 & 0.82 & 0.15 & 0.33 \\
$\lambda_{\text{xyz}}{=}10^4$ & 1.46 & 0.27 & 0.10 & 0.85 & 0.17 & 0.39 \\
$\lambda_{\text{xyz}}{=}10^6$ & 0.03 & 0.26 & 0.10 & 0.83 & 0.16 & 0.38 \\
Ours & 0.24 & $\mathbf{0.28}$ & $\mathbf{0.11}$ & $\mathbf{0.89}$ & $\mathbf{0.19}$ &  $\mathbf{0.41}$ \\

\bottomrule
\end{tabular}
}
\label{tab:ablations_drift}
\end{table}

\smallskip
\noindent \textbf{Physics Prior Robustness.}
We evaluate the robustness of our framework to perturbations in key simulation parameters. We perturb four parameters by $\pm$50\%: emission rate, initial velocity, Young's modulus, and movement threshold $\delta$. Experiments are conducted on all scenes across snowfall, rainfall, and sandstorm effects. As shown in Tab.~\ref{tab:robustness}, our method maintains stable performance across all perturbations and motion remains stable with average trajectory drift (Eq.~\ref{eq:drift_eq}) of $0.28\pm0.03$ compared to a baseline drift of $0.26$ on these three effects.

\begin{table}[t]
\centering
\caption{\textbf{Physics Prior Robustness}. Performance under $\pm$50\% perturbation of key simulation parameters (emission rate, initial velocity, Young's modulus, movement threshold), averaged across all scenes and three effects (snowfall, rainfall, sandstorm). Our method maintains stable performance across all perturbations.}
\resizebox{0.99\linewidth}{!}{\begin{tabular}{lccccc}
\toprule
Setting & $\text{CLIP}_{Sim}\uparrow$ & $\text{CLIP}_{Dir} \uparrow$ & $\text{VQAScore}\uparrow$ & $\text{ViCLIP-T}\uparrow$ & $\text{VE-Bench}\uparrow$ \\
\midrule
$\pm$50\% Perturbation & $0.27{\pm}0.02$ & $0.10{\pm}0.01$ & $0.87{\pm}0.04$ & $0.17{\pm}0.01$ & $0.38{\pm}0.03$ \\
Default (Ours) & $\mathbf{0.28}$ & $\mathbf{0.11}$ & $\mathbf{0.88}$ & $\mathbf{0.19}$ & $\mathbf{0.39}$ \\
\bottomrule
\end{tabular}
}
\label{tab:robustness}
\end{table}

\smallskip
\noindent \textbf{Regularization Weight Sensitivity.}
We vary each regularization weight independently by $\times 0.1$ and $\times 10$ while keeping others at their default values. Experiments are conducted on all scenes across snowfall and sandstorm effects. As shown in Tab.~\ref{tab:weight_sensitivity}, physics regularization weights ($\lambda_{\text{xyz}}$, $\lambda_{\text{vel}}$, $\lambda_{\text{rot}}$) are robust to order-of-magnitude changes, with metrics remaining within 0.01-0.02 of default values. Appearance regularization weights are more sensitive when reduced by $10\times$ but degrade gracefully when increased, confirming a stable operating range around our defaults.

\begin{table}[t]
\centering
\caption{\textbf{Regularization Weight Sensitivity}. Each weight is varied independently by $\times 0.1$ and $\times 10$ on all scenes across snowfall and sandstorm effects. Physics weights are robust to order-of-magnitude changes. Appearance weights are sensitive when reduced but degrade gracefully when increased.}
\resizebox{0.99\linewidth}{!}{\begin{tabular}{lcccccc}
\toprule
Weight & Value & $\text{CLIP}_{Sim}\uparrow$ & $\text{CLIP}_{Dir} \uparrow$ & $\text{VQAScore}\uparrow$ & $\text{ViCLIP-T}\uparrow$ & $\text{VE-Bench}\uparrow$ \\
\midrule
\multirow{2}{*}{$\lambda_{\text{xyz}}$ (0.1)} & $\times 0.1$ & 0.26 & 0.09 & 0.84 & 0.16 & 0.35 \\
& $\times 10$ & 0.27 & 0.10 & 0.86 & 0.17 & 0.37 \\
\midrule
\multirow{2}{*}{$\lambda_{\text{vel}}$ (0.1)} & $\times 0.1$ & 0.26 & 0.10 & 0.85 & 0.16 & 0.37 \\
& $\times 10$ & 0.27 & 0.10 & 0.86 & 0.17 & 0.39 \\
\midrule
\multirow{2}{*}{$\lambda_{\text{rot}}$ (0.1)} & $\times 0.1$ & 0.28 & $\mathbf{0.11}$ & 0.88 & $\mathbf{0.19}$ & 0.39 \\
& $\times 10$ & 0.28 & $\mathbf{0.11}$ & 0.88 & $\mathbf{0.19}$ & 0.41 \\
\midrule
\multirow{2}{*}{$\lambda^{\text{active}}_{\sigma}$ (1.0)} & $\times 0.1$ & 0.24 & 0.08 & 0.79 & 0.15 & 0.32 \\
& $\times 10$ & 0.27 & 0.10 & 0.86 & 0.18 & 0.39 \\
\midrule
\multirow{2}{*}{$\lambda^{\text{active}}_{\mathbf{S}}$ (1.0)} & $\times 0.1$ & 0.19 & 0.04 & 0.52 & 0.09 & 0.19 \\
& $\times 10$ & 0.26 & 0.10 & 0.83 & 0.17 & 0.37 \\
\midrule
\multirow{2}{*}{$\lambda^{\text{active}}_{\mathbf{C}}$ (35.0)} & $\times 0.1$ & 0.20 & 0.06 & 0.57 & 0.11 & 0.23 \\
& $\times 10$ & 0.26 & 0.10 & 0.84 & 0.18 & 0.37 \\
\midrule
\multirow{2}{*}{$\lambda^{\text{collided}}_{\sigma}$ (35.0)} & $\times 0.1$ & 0.22 & 0.07 & 0.74 & 0.13 & 0.30 \\
& $\times 10$ & 0.27 & 0.10 & 0.85 & 0.18 & 0.38 \\
\midrule
\multirow{2}{*}{$\lambda^{\text{collided}}_{\mathbf{S}}$ (35.0)} & $\times 0.1$ & 0.17 & 0.03 & 0.47 & 0.07 & 0.14 \\
& $\times 10$ & 0.26 & 0.10 & 0.84 & 0.18 & 0.38 \\
\midrule
\multirow{2}{*}{$\lambda^{\text{collided}}_{\mathbf{C}}$ (35.0)} & $\times 0.1$ & 0.21 & 0.06 & 0.61 & 0.12 & 0.26 \\
& $\times 10$ & 0.27 & 0.10 & 0.85 & 0.18 & 0.37 \\
\midrule
\multicolumn{2}{l}{Default (Ours)} & $\mathbf{0.29}$ & $\mathbf{0.11}$ & $\mathbf{0.89}$ & $\mathbf{0.19}$ & $\mathbf{0.42}$ \\
\bottomrule
\end{tabular}
}
\label{tab:weight_sensitivity}
\end{table}

\smallskip
\noindent \textbf{Scheduler Comparison.}
We compare our SDS-adaptive scaling against three fixed regularization weight decay schedules on all scenes across snowfall and sandstorm effects: (1) linear decay ($\lambda$: $10^6 \rightarrow 10^2$ over 1000 iterations), (2) exponential decay ($\lambda{=}10^6 \cdot 0.995^{\text{iter}}$), and (3) cosine annealing ($\lambda$: $10^6 \rightarrow 10^2$ following a cosine curve). As shown in Tab.~\ref{tab:schedulers}, our SDS-adaptive scaling outperforms all fixed schedules. Fixed schedules fail because Video-SDS loss magnitude varies unpredictably during training, while our adaptive scaling strengthens physics guidance when SDS is uncertain and relaxes it when the model is confident.

\begin{table}[t]
\centering
\caption{\textbf{Scheduler Comparison}. Comparison of fixed regularization weight decay schedules versus our SDS-adaptive scaling on all scenes across snowfall and sandstorm effects. Our adaptive approach outperforms all fixed schedules.}
\resizebox{0.99\linewidth}{!}{\begin{tabular}{lccccc}
\toprule
Schedule & $\text{CLIP}_{Sim}\uparrow$ & $\text{CLIP}_{Dir} \uparrow$ & $\text{VQAScore}\uparrow$ & $\text{ViCLIP-T}\uparrow$ & $\text{VE-Bench}\uparrow$ \\
\midrule
Linear Decay & 0.25 & 0.09 & 0.82 & 0.15 & 0.35 \\
Exponential Decay & 0.24 & 0.08 & 0.80 & 0.14 & 0.34 \\
Cosine Annealing & 0.26 & 0.10 & 0.84 & 0.17 & 0.38 \\
Ours (SDS-Adaptive) & $\mathbf{0.29}$ & $\mathbf{0.11}$ & $\mathbf{0.89}$ & $\mathbf{0.19}$ & $\mathbf{0.42}$ \\
\bottomrule
\end{tabular}
}
\label{tab:schedulers}
\end{table}

\smallskip
\noindent \textbf{Per-Particle Appearance and Rotation Ablation.}
We ablate the necessity of per-particle appearance variation and angular velocity prediction. We evaluate two variants on all scenes across snowfall and sandstorm effects: (1) \emph{fixed appearance}, where all active particles share one optimized appearance and all collided particles share another, and (2) \emph{fixed rotation}, where the angular velocity prediction is disabled. The variance of learned parameters across particles, averaged over timesteps, scenes, and effects, confirms meaningful per-particle variation for appearance: $\text{Var}(\sigma)_{\text{active}}{=}0.12$, $\text{Var}(\mathbf{S})_{\text{active}}{=}0.21$, $\text{Var}(\mathbf{C})_{\text{active}}{=}0.05$, $\text{Var}(\sigma)_{\text{collided}}{=}0.01$, $\text{Var}(\mathbf{S})_{\text{collided}}{=}0.03$, $\text{Var}(\mathbf{C})_{\text{collided}}{=}0.005$, while angular velocity variance is near-zero ($\text{Var}(\omega){\approx}0$).

Tab.~\ref{tab:per_particle} demonstrates that removing per-particle appearance clearly degrades performance across all metrics, confirming its importance. Fixed rotation shows no degradation, consistent with the near-zero angular velocity variance ($\text{Var}(\omega){\approx}0$). We retain this capacity for generality and for effects where rotation may be more prominent.

\begin{table}[t]
\centering
\caption{\textbf{Per-Particle Appearance and Rotation Ablation}. Removing per-particle appearance degrades all metrics. Removing angular velocity shows no degradation, consistent with near-zero learned rotation variance for weather effects.}
\resizebox{0.99\linewidth}{!}{\begin{tabular}{lccccc}
\toprule
Variant & $\text{CLIP}_{Sim}\uparrow$ & $\text{CLIP}_{Dir} \uparrow$ & $\text{VQAScore}\uparrow$ & $\text{ViCLIP-T}\uparrow$ & $\text{VE-Bench}\uparrow$ \\
\midrule
Fixed Appearance & 0.26 & 0.09 & 0.83 & 0.17 & 0.36 \\
Fixed Rotation & $\mathbf{0.29}$ & $\mathbf{0.11}$ & 0.88 & $\mathbf{0.19}$ & $\mathbf{0.43}$ \\
Ours & $\mathbf{0.29}$ & $\mathbf{0.11}$ & $\mathbf{0.89}$ & $\mathbf{0.19}$ & 0.42 \\
\bottomrule
\end{tabular}
}
\label{tab:per_particle}
\end{table}

\smallskip
\noindent \textbf{SDS Variant Comparison.}
We evaluate two improved score distillation variants to assess compatibility with our framework, Noise-Free Score Distillation (NFSD)~\cite{katzir2023noise}, which decomposes the score into interpretable components and removes an undesired noise term to reduce over-smoothing, and Bridge-SDS~\cite{mcallister2024rethinking}, which formulates score distillation as optimal transport between source and target distributions to reduce oversaturation artifacts. We replace our vanilla SDS objective with each variant and evaluate on all scenes across snowfall and sandstorm effects, keeping all other components identical.

As shown in Tab.~\ref{tab:sds_variants}, both improved variants yield higher scores across all metrics, with Bridge-SDS achieving the strongest results. While our physics-guided optimization drives the primary quality improvements over baselines (Tab.~\ref{tab:comparisons}), these results confirm that orthogonal advances in score distillation provide additional gains when integrated into our framework.

\begin{table}[t]
\centering
\caption{\textbf{SDS Variant Comparison}. Our framework is compatible with improved score distillation methods. Both NFSD~\cite{katzir2023noise} and Bridge-SDS~\cite{mcallister2024rethinking} improve over vanilla SDS when integrated into our physics-guided optimization.}
\resizebox{0.99\linewidth}{!}{\begin{tabular}{lccccc}
\toprule
SDS Variant & $\text{CLIP}_{Sim}\uparrow$ & $\text{CLIP}_{Dir} \uparrow$ & $\text{VQAScore}\uparrow$ & $\text{ViCLIP-T}\uparrow$ & $\text{VE-Bench}\uparrow$ \\
\midrule
Vanilla SDS & 0.29 & 0.11 & 0.89 & 0.19 & 0.42 \\
NFSD & 0.30 & 0.11 & 0.91 & 0.20 & 0.44 \\
Bridge-SDS & $\mathbf{0.32}$ & $\mathbf{0.12}$ & $\mathbf{0.93}$ & $\mathbf{0.22}$ & $\mathbf{0.45}$ \\
\bottomrule
\end{tabular}
}
\label{tab:sds_variants}
\end{table}

\begin{figure}[t]
    \centering
    \footnotesize
    \setlength\tabcolsep{0.05em}
    \begin{tabular}{ccc}
    \rotatebox{90}{\whitetxt{xxxxxxxx}Blender} &
    \includegraphics[width=0.48\linewidth]{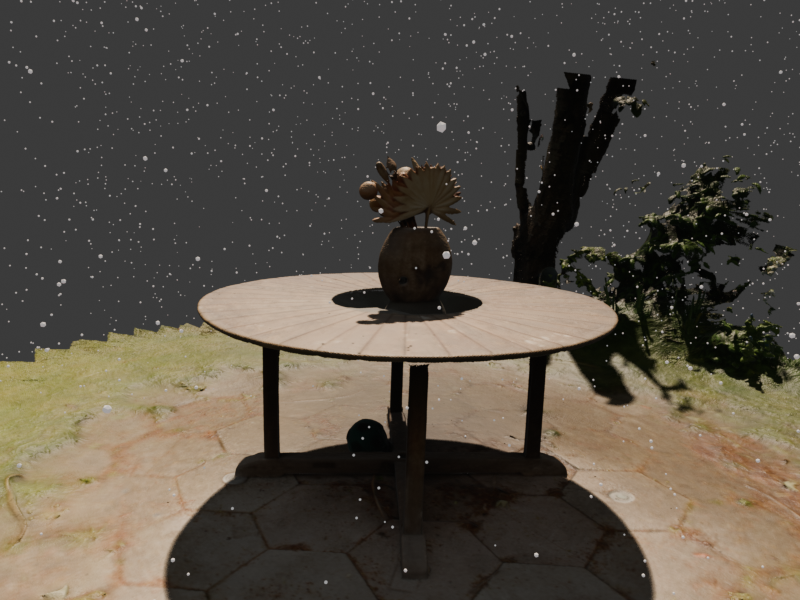} &
    \includegraphics[width=0.48\linewidth]{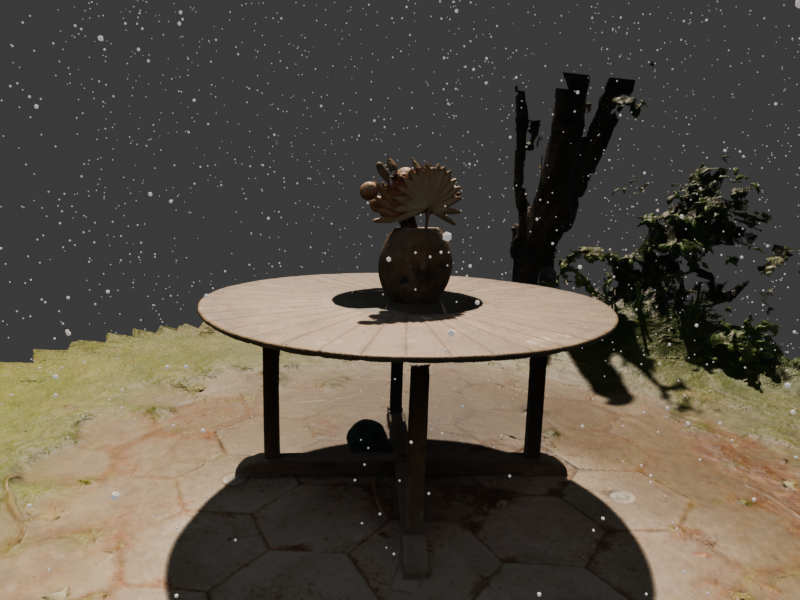} \\
    \rotatebox{90}{\whitetxt{xxxxxxxx}Ours} &
    \includegraphics[width=0.48\linewidth]{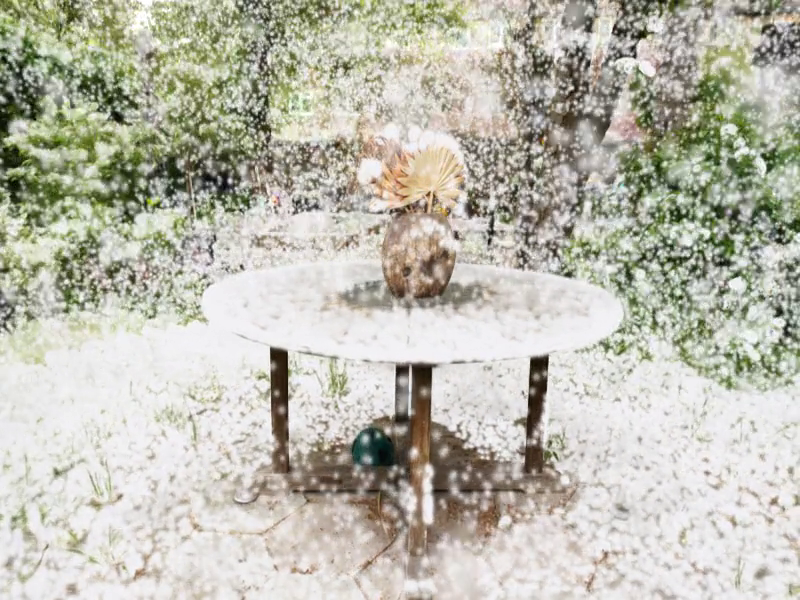} &
    \includegraphics[width=0.48\linewidth]{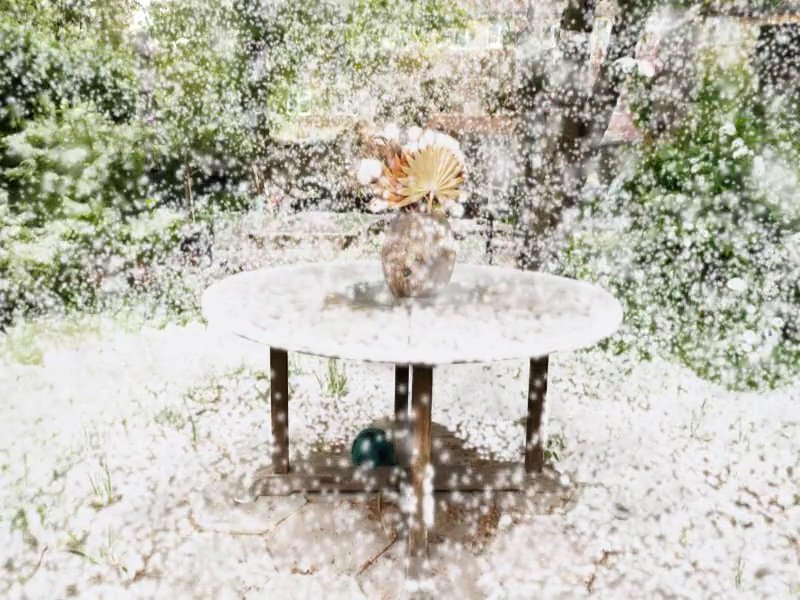} \\
    \end{tabular}
    \vspace{-2mm}
    \caption{\textbf{Comparison to Manual VFX}. Two consecutive frames of a snowfall effect on the Garden scene created manually in Blender~\cite{blender} (${\sim}$10 hours, top) versus our automated pipeline (${\sim}$1 hour, bottom). Our automated pipeline produces results of comparable visual quality while requiring an order of magnitude less time and no manual expertise.}
    \label{fig:blender_comp}
\end{figure}

\smallskip
\noindent \textbf{Comparison to Manual VFX.}
To contextualize the quality-automation tradeoff, we created an equivalent snowfall effect in Blender~\cite{blender} on the Garden scene. As a novice user, manual setup in Blender required approximately 10 hours (particle system configuration, material setup, lighting adjustment, rendering), compared to our automated pipeline of approximately 1 hour. A visual comparison is provided in Fig.~\ref{fig:blender_comp}. Our results are comparable or better than the manually created Blender effect, suggesting that little to no realism is lost for the sake of automation while requiring significantly less time and expertise.

\section{Additional Details}
\label{sec:supp_details}

\subsection{Implementation Details}
\label{sec:imp_details}

For each scene we first use 3D Gaussian Splatting \cite{kerbl20233d} to train our static 3D representation, using the default hyper parameters found in the code provided by the authors (\href{https://github.com/graphdeco-inria/gaussian-splatting}{https://github.com/graphdeco-inria/gaussian-splatting}). To extract meshes we follow the mesh extraction method in \cite{huang20242d} by using the code provided by the authors (\href{https://github.com/hbb1/2d-gaussian-splatting/}{https://github.com/hbb1/2d-gaussian-splatting/}). We render depth maps of the training views using the depth value of the splats projected to the pixels and utilize truncated signed distance fusion (TSDF) to fuse the reconstruction depth maps, using Open3D~\cite{Zhou2018}. We set the voxel size to $0.004$, the truncated threshold to 0.02 during TSDF fusion and extract a mesh with resolution of 1024.

For the simulation, we base our MPM simulation on Taichi~\cite{hu2019taichi}, which supports various types of materials (e.g., \textit{snow}, \textit{water}, \textit{sand}, etc.). We build on \cite{qiu2024feature}, using the code provided by the authors (\href{https://github.com/vuer-ai/feature-splatting-inria}{https://github.com/vuer-ai/feature-splatting-inria}). We first align the Gaussians to the particle system coordinates and then normalize them to fit in a unit cube and center them appropriately with the bottom aligned to a ground plane. We then add the static particles to the simulation as ``stationary" material particles.

We then add particles based on their effect-specific motion simulation parameters (detailed in Sec. \ref{sec:simulation_params}). We run the simulation for 500 steps.
We set the threshold for particle active state updates at 0.001. Our grid resolution is set to $64^3$. After each simulation step we transform the positions back to the Gaussian world and save them for rendering. We also save the collided Gaussians position at the time of collision. 

For physics-guided score distillation optimization, we implement separate neural dynamics models for active and collided Gaussian states. Each model consists of a shared MLP backbone with separate prediction heads. The models take as input: (1) XYZ positions encoded through Fourier feature embeddings (24 dimensions), (2) physics velocities encoded similarly (24 dimensions), (3) time encoded through sinusoidal embeddings (24 dimensions), and (4) the previous state's appearance parameters. The shared backbone consists of two hidden layers with 128 hidden dimensions and ReLU activations. For active Gaussians, the model predicts velocity corrections $\Delta\mathbf{v}$, angular velocities $\boldsymbol{\omega}$, and appearance deltas ($\Delta\sigma$, $\Delta\mathbf{S}$, $\Delta\mathbf{C}$). For collided Gaussians, the model predicts only appearance deltas as their motion is fixed at collision states.

We employ Video Score Distillation Sampling using CogVideoX-2B~\cite{yang2024cogvideox} as the video diffusion model. During optimization, we render video clips at $720\times480$ resolution with a length of 9 frames from randomly sampled training camera viewpoints. We use a guidance scale of 100 and optimize for 1000 iterations with an initial learning rate of $1\times10^{-4}$, using a cosine annealing schedule. For timestep sampling, we employ progressive annealing, sampling timesteps from $[t_{\text{min}}, t_{\text{max}}]$, where $t_{\text{min}} = 20$ and $t_{\text{max}}$ decreases linearly from 980 to 740. To balance the gradients between the diffusion prior and regularization terms, we employ SDS-adaptive physics guidance where all regularization weights are dynamically scaled by the instantaneous magnitude of the Video-SDS loss ($|\mathcal{L}_{\text{SDS}}|$).  Consequently, the regularization weights act as a percentage relative to the diffusion gradient magnitude. We set the physics regularization weights as follows: $\lambda_{\text{xyz}} = 0.1$, $\lambda_{\text{vel}} = 0.1$, $\lambda_{\text{rot}} = 0.1$. We set the appearance regularization weights as follows: for active particles, $\lambda^{\text{active}}_{\sigma} = 1.0$, $\lambda^{\text{active}}_{\mathbf{S}} = 1.0$, and $\lambda^{\text{active}}_{\mathbf{C}} = 35.0$; for collided particles, we enforce stronger regularization with $\lambda^{\text{collided}}_{\sigma} = 35.0$, $\lambda^{\text{collided}}_{\mathbf{S}} = 35.0$, and $\lambda^{\text{collided}}_{\mathbf{C}} = 35.0$.

\smallskip
\noindent \textbf{Runtime.}
Our framework runs on a single NVIDIA A100 40GB GPU. A runtime comparison against baselines is provided in Tab.~\ref{tab:runtime}.
\begin{table}[h]
\centering
\caption{\textbf{Runtime Comparison}. Per scene-effect runtime on a single NVIDIA A100 40GB GPU.}
\small
\begin{tabular}{@{}lr@{}}
\toprule
Method & Runtime \\
\midrule
ClimateNeRF~\cite{li2023climatenerf} & ${\sim}$180 min \\
GaussCtrl~\cite{wu2024gaussctrl} & ${\sim}$30 min \\
Instruct-4DGS~\cite{kwon2025efficient} & ${\sim}$45 min \\
Ours & ${\sim}$60 min \\
\bottomrule
\end{tabular}
\label{tab:runtime}
\end{table}

\subsection{Collision Handling Details}
\label{sec:ch_details}
Following Sec.~\ref{sec:motion_stage}, we describe our mesh-based collision handling.

\noindent \textbf{\textit{Snowfall} Collision Handling}. 
For the \textit{snowfall} effect we employ a dual approach. First, for surface accumulation, we compute the closest surface point and surface normal on the mesh to each interacting Gaussian using Open3D~\cite{Zhou2018}. We project the Gaussians onto their closest surface point along the surface normal.
Second, for background enhancement during the final rendering, we finetune our pretrained 3DGS model on images generated by a pretrained ClimateNeRF~\cite{li2023climatenerf} model for snow. This finetuning process runs for 1000 iterations, optimizing only the scale, opacity, and color parameters of the background Gaussians (defined as all Gaussians outside our extracted mesh bounds). We then interpolate these parameters over the timesteps to create a gradual snow effect in the background environment.

\smallskip
\noindent \textbf{\textit{Sandstorm} Collision Handling}.  
For the \textit{sandstorm} effect we perform the same surface projection as in the \textit{snowfall} effect. We also use anisotropic scaling on the accumulated Gaussians in order to create thin, flat accumulations.

\smallskip
\noindent \textbf{\textit{Rainfall} Collision Handling}.  
For the \textit{rainfall} effect, we also employ a dual approach. First, for surface accumulation, we perform the same surface projection as in the \textit{snowfall} effect, but instead of accumulating visible Gaussians, we set these collided Gaussians' opacity to zero.
Second, we utilize a volumetric grid-based approach to model surface wetness for the final rendering. When rain Gaussians collide with the mesh, we compute the closest surface points and increase wetness values in a coarse 3D grid surrounding the mesh, applying a Gaussian smoothing kernel to spread the effect naturally around impact points. The wetness grid undergoes temporal decay, simulating the gradual drying of surfaces. During rendering, we modify the appearance of the original Gaussians based on local wetness values, darkening affected areas to create realistic wet surface effects.

\subsection{Motion Simulation Parameters}
\label{sec:simulation_params}

Following Sec.~\ref{sec:motion_stage}, we detail our motion simulation parameters.

\smallskip
\noindent \textbf{\textit{Snowfall}.}  
Our \textit{snowfall} simulation uses a Young's modulus of 0.14 and Poisson ratio of 0.2. We emit 1000 particles every 2 frames from random positions above the scene with initial vertical velocity of -0.5 and horizontal variation between -0.1 and 0.1. Standard gravity (0, -9.8, 0) is applied. 

\smallskip
\noindent \textbf{\textit{Rainfall}.}  
Our \textit{rainfall} simulation uses a Young's modulus of 0.08 and Poisson ratio of 0.45. We emit 1000 particles every 2 frames from random positions above the scene with initial vertical velocity of -0.5 and horizontal variation between -0.1 and 0.1. Standard gravity (0, -9.8, 0) is applied.

\smallskip
\noindent \textbf{\textit{Sandstorm}.}  
Our \textit{sandstorm} simulation uses a Young's modulus of 0.08 and Poisson ratio of 0.3. We emit 1000 particles every 2 frames from one side of the scene with strong horizontal velocity (0.8-1.2) and minor vertical variation (-0.2 to 0.2). Standard gravity (0, -9.8, 0) is applied.

\smallskip
\noindent \textbf{\textit{Fog}.}   
Our \textit{fog} simulation differs by emitting 5000 water particles throughout the scene volume rather than from above. Particles have minimal initial velocity variation (-0.1 to 0.1) with slight horizontal drift (0.5 in x-direction). Nearly negligible gravity (0.5, -0.1, 0) is applied.

\subsection{Appearance Initialization}
\label{sec:appearance_init}
Following Sec.~\ref{sec:appearance_stage}, we detail our appearance parameter LLM-initialization. We employ a large language model (Claude Sonnet 4, Anthropic, 2025) to generate baseline parameters $A^{active}_{init}$ and $A^{collided}_{init}$ for each effect. We prompt the LLM with detailed descriptions to obtain appropriate scale, opacity, and color parameters. 

For example, for snowfall we use: 
\begin{verbatim}
"Generate appearance parameters for 
realistic falling snow particles in a 
3D Gaussian Splatting scene. 
Provide: scale, opacity, and RGB color 
values. Format as numerical values."
\end{verbatim}

We then optimize these initialized parameters using Video-SDS with the following text prompts combining effect descriptions with scene-specific details:

\smallskip
\noindent\textbf{Optimization Prompts:}
\begin{itemize}
    \item \emph{Snowfall}: "Fluffy snowflakes are falling in a [SCENE DESCRIPTION], accumulating on surfaces. Photorealistic, high detail"
    \item \emph{Rainfall}: "Rain falling in a [SCENE DESCRIPTION], wettening the surfaces. Photorealistic, high detail"
    \item \emph{Sandstorm}: "A sandstorm engulfs the [SCENE DESCRIPTION], with swirling clouds of sand in the air and the sand accumulates on surfaces. Photorealistic, high detail"
    \item \emph{Fog}: "Fog moving across the [SCENE DESCRIPTION], Photorealistic, high detail"
\end{itemize}

\smallskip
\noindent\textbf{Scene Descriptions:}
\begin{itemize}
    \item \emph{Bicycle}: "A bicycle leaning on a bench in the park"
    \item \emph{Garden}: "A table with a vase on it in a garden"
    \item \emph{Stump}: "An old, hollow tree stump covered in ivy lying on the forest floor"
    \item \emph{Playground}: "A playground with slides and benches around it"
    \item \emph{Truck}: "An old, light blue truck parked on a city street"
\end{itemize}

\subsection{Comparisons and Ablation Details}
\label{sec:comp_details}

Below we provide all the details needed to reproduce the comparisons and ablations shown in the paper.

\subsubsection*{Baseline Methods}

\smallskip
\noindent \textbf{ClimateNeRF.}~\cite{li2023climatenerf} 
We use the code provided by the authors (\href{https://github.com/y-u-a-n-l-i/Climate_NeRF}{https://github.com/y-u-a-n-l-i/Climate\_NeRF}) with the default parameters.

\smallskip
\noindent \textbf{GaussCtrl.}~\cite{wu2024gaussctrl} 
We use the code provided by the authors (\href{https://github.com/ActiveVisionLab/gaussctrl}{https://github.com/ActiveVisionLab/gaussctrl}) with the default parameters.

\subsubsection*{Ablation Details}

\smallskip
\noindent \textbf{w/o Collision Handling \emph{(w/o CH)}.} 
We disable the mesh-based collision handling techniques described in Sec. \ref{sec:ch_details}. Dynamic Gaussians follow their simulated trajectories without surface projection or accumulation adjustments, resulting in Gaussians floating above surfaces rather than realistically interacting with scene geometry.

\smallskip
\noindent \textbf{w/o Appearance Optimization \emph{(w/o App)}.} 
We skip the Video-SDS appearance optimization stage entirely and directly use the LLM-initialized parameters described in Sec. \ref{sec:appearance_init}. This variant uses only the baseline appearance parameters $\mathcal{A}^{active}_{init}$ and $\mathcal{A}^{collided}_{init}$ without any refinement through diffusion guidance.

\smallskip
\noindent \textbf{w/o Motion Simulation \emph{(w/o Motion)}.} 
We bypass our physics-based motion simulation and instead use 4D Gaussian Splatting~\cite{wu20244d}, with default parameters as in the code provided by the authors (\href{https://github.com/hustvl/4DGaussians}{https://github.com/hustvl/4DGaussians}). We initialize a deformation field and optimize it directly using Video-SDS with the same prompts and optimization settings as our full method. This variant attempts to learn both motion and appearance simultaneously from the video diffusion model alone.

\smallskip
\noindent \textbf{w/o Physics Guidance \emph{(w/o PG)}.} 
We fix the motion trajectories from physics simulation and optimize only the appearance matrices ($\mathcal{A}^{\text{active}}$, $\mathcal{A}^{\text{collided}}$) directly through Video-SDS. This variant eliminates both physics conditioning (the neural dynamics model does not receive velocities as input) and physics regularization losses ($\mathcal{L}_{\text{xyz}}$, $\mathcal{L}_{\text{vel}}$, $\mathcal{L}_{\text{rot}}$). The model predicts only appearance deltas ($\boldsymbol{\Delta}\mathcal{A}$) and does not predict velocity corrections ($\boldsymbol{\Delta}\mathbf{v}$) or angular velocities ($\boldsymbol{\omega}$), preventing joint optimization of motion and appearance. We retain only appearance regularization ($\mathcal{L}_{\text{app}}$) to prevent error accumulation in the recurrent predictions.

\subsubsection*{Evaluation Protocol}

For quantitative evaluation of video-based metrics, we generate 15 camera trajectories per scene by interpolating between randomly sampled pairs of training viewpoints. We render videos along these trajectories for all methods and compute the metrics (ViCLIP-T, VE-Bench) on the resulting videos. For image-based metrics, we render  images from all training viewpoints.

\section{Additional Visualizations and Results}
\label{sec:supp_vis}

\begin{figure*}
  \centering
  \includegraphics[width=\textwidth, trim={3.5cm 17.5cm 4.7cm 3.2cm},clip]{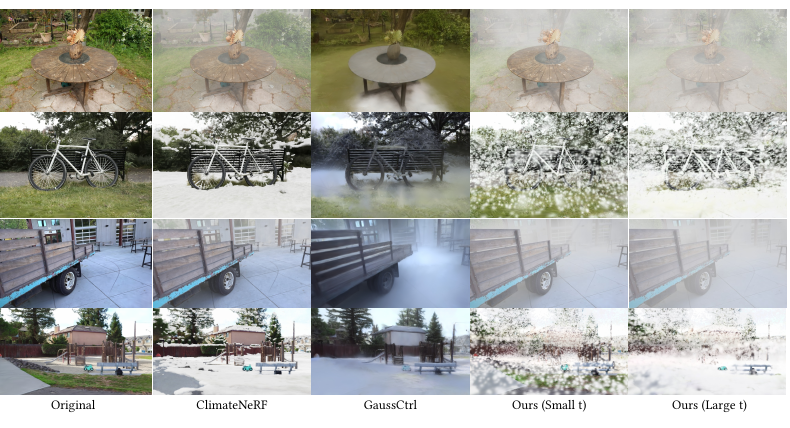}
  \vspace{-0.8cm}
  \caption{\textbf{Additional comparisons to static 3D editing techniques}. We show additional qualitative comparisons between static 3D editing methods and our dynamic approach.  Only the \textit{fog} (shown on the top row and the third row form the top) and \textit{snowfall} (shown on the bottom row and the second row from the top) effects are shown, because ClimateNeRF only supports editing for these effects. We compare the original scene, ClimateNeRF~\cite{li2023climatenerf}, GaussCtrl~\cite{wu2024gaussctrl}, and our method at both early (Small t) and later (Large t) simulation timesteps.}
\label{fig:supp_comp}
\end{figure*}

\begin{figure*}
  \centering
  \includegraphics[width=\textwidth, trim={3.5cm 15.0cm 5.0cm 3.2cm},clip]{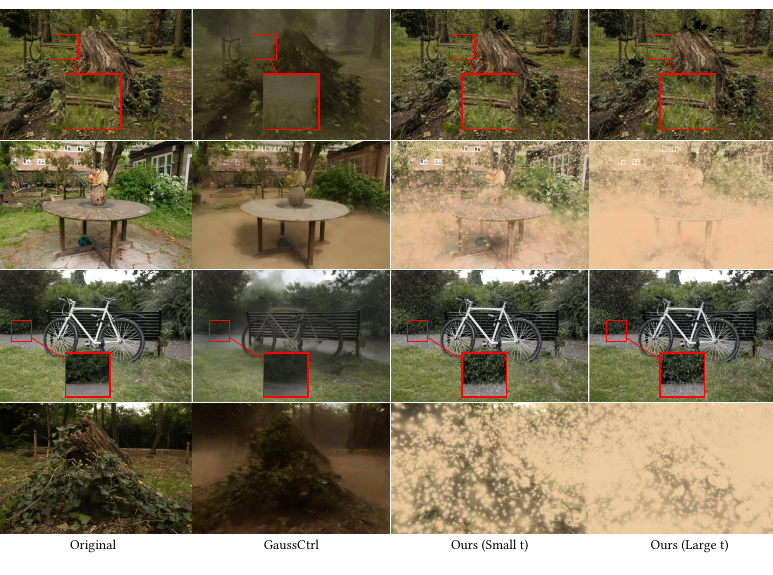}
  \vspace{-1.0cm}
  \caption{\textbf{Additional Comparisons on the \textit{sandstorm} and \textit{rainfall} edits}. We show additional comparisons between our approach and GaussCtrl~\cite{wu2024gaussctrl}. We show results for the \textit{sandstorm} (shown on the bottom row and the second row from the top) and \textit{rainfall}  (shown on the top row and the second row from the bottom) effects. We compare the original scene,  GaussCtrl~\cite{wu2024gaussctrl}, and our method at both early (Small t) and later (Large t) simulation timesteps. We note that ClimateNeRF~\cite{li2023climatenerf} does not support editing for these effects. We zoom in on the results for \textit{rainfall}, as these results are challenging to visualize.}
\label{fig:supp_comp2}
\end{figure*}

We present additional comparisons to the static 3D editing methods ClimateNeRF~\cite{li2023climatenerf} and GaussCtrl~\cite{wu2024gaussctrl} in Figure \ref{fig:supp_comp}.
We also show comparisons to GaussCtrl over the \textit{sandstorm} and \textit{rainfall} effects in Figure \ref{fig:supp_comp2}.

\end{document}